\documentclass{aa}
\usepackage{graphicx}
\usepackage{hyperref}

\usepackage{natbib}
\bibpunct{(}{)}{;}{a}{}{,} 

\usepackage{txfonts}
\usepackage{subfig}

\begin{document}

\title{ Painting a portrait of the Galactic disc with its stellar clusters \thanks{List of cluster parameters and complete list of their members available in electronic form only at the CDS via anonymous ftp to cdsarc.u-strasbg.fr(130.79.128.5) or via http://cdsweb.u-strasbg.fr/cgi-bin/qcat?J/A+A/} }

   \author{T. Cantat-Gaudin\inst{\ref{UB}}
   \and
F. Anders\inst{\ref{UB}}
\and
A. Castro-Ginard\inst{\ref{UB}}
\and
C. Jordi\inst{\ref{UB}}
\and
M. Romero-G{\'o}mez\inst{\ref{UB}}
\and
C. Soubiran\inst{\ref{LAB}}
\and
L. Casamiquela\inst{\ref{LAB}}
\and
Y. Tarricq\inst{\ref{LAB}}
\and
A. Moitinho\inst{\ref{CENTRA}}
\and
A. Vallenari\inst{\ref{OAPD}}
\and
A. Bragaglia\inst{\ref{OABO}}
\and
A. Krone-Martins\inst{\ref{CENTRA},\ref{UCI}}
\and
M. Kounkel\inst{\ref{WA}}
        }

  \institute{
                Institut de Ci\`encies del Cosmos, Universitat de Barcelona (IEEC-UB), Mart\'i i Franqu\`es 1, E-08028 Barcelona, Spain\label{UB}
                \\
                \email{tcantat@fqa.ub.edu}
                \and
Laboratoire d’Astrophysique de Bordeaux, Univ. Bordeaux, CNRS, UMR 5804, 33615 Pessac, France\label{LAB}
\and
CENTRA, Faculdade de Ci\^encias, Universidade de Lisboa, Ed. C8, Campo Grande, 1749-016 Lisboa, Portugal\label{CENTRA}
\and
INAF-Osservatorio Astronomico di Padova, vicolo Osservatorio 5, 35122 Padova, Italy\label{OAPD}
\and
INAF-Osservatorio di Astrofisica e Scienza dello Spazio, via Gobetti 93/3, 40129 Bologna, Italy\label{OABO}
 \and Donald Bren School of Information and Computer Sciences,
University of California, Irvine, CA 92697, USA\label{UCI}
\and
Department of Physics and Astronomy, Western Washington University, 516 High St, Bellingham, WA 98225, USA\label{WA}
  }

   \date{Received ???, ???; accepted ???, ???}

  \abstract{The large astrometric and photometric survey performed by the \textit{Gaia} mission allows for a panoptic view of the Galactic disc and in its stellar cluster population. Hundreds of stellar clusters were only discovered after the latest \textit{Gaia} data release (DR2) and have yet to be characterised.}{Here we make use of the deep and homogeneous \textit{Gaia} photometry down to $G$=18 to estimate the distance, age, and interstellar reddening for about 2000 stellar clusters identified with \textit{Gaia}~DR2 astrometry. We use these objects to study the structure and evolution of the Galactic disc.}{ We relied on a set of objects with well-determined parameters in the literature to train an artificial neural network to estimate parameters from the \textit{Gaia} photometry of cluster members and their mean parallax.}{We obtain reliable parameters for 1867 clusters. Our catalogue confirms the relative lack of old stellar clusters in the inner disc (with a few notable exceptions). We also quantify and discuss the variation of scale height with cluster age, and we detect the Galactic warp in the distribution of old clusters.}{This work results in a large and homogeneous cluster catalogue, allowing one to trace the structure of the disc out to distances of $\sim$4\,kpc. However, the present sample is still unable to trace the outer spiral arm of the Milky Way, which indicates that the outer disc cluster census might still be incomplete.}

\titlerunning{   Painting a portrait of the Galactic disc with its stellar clusters    }
\authorrunning{T. Cantat-Gaudin et al.}
\maketitle


\section{Introduction}

The shape and dimension of our galaxy, which we commonly refer to as the Milky Way, is difficult to appreciate from our vantage point. 
From the pioneering work of early modern astronomers \citep{Herschel1785,Shapley18,Trumpler30} to recent studies \citep{Reid19,Gravity19,Anders19}, the distance to individual objects is one of the most valuable pieces of information we rely on to reconstruct the overall structure of the Milky Way.

Among the variety of astronomical objects to which we can derive distances, stellar clusters present the advantage of spanning a wide range of ages, from a few million years (tracing episodes of recent star formation) to several gigayears (as old as the Galactic disc itself), which can be estimated with a greater precision than for individual stars. 
Samples of clusters with known ages have long been used to trace various properties of the Galactic disc, such as the path of its spiral arms \citep{Becker70} or the evolution of its scale height \citep{vandenBergh58}.
Although the precision and accuracy of age estimates are tied to the quality of the observational data and the correctness of theoretical models, distinguishing a young cluster from an old one is often relatively straightforward in a colour-magnitude diagram\footnote{\citet{Trumpler25} was the first to group clusters by age according to their magnitude-spectral class diagrams, but his evolutionary sequence was wrong. It was then believed that stars formed as giants and contracted into main-sequence dwarfs \citep[see][for a discussion]{Sandage88}.}. While the first catalogues of cluster parameters only reported sky coordinates \citep[e.g.][]{Melotte15} and sometimes distances \citep{Trumpler30,Collinder31}, modern catalogues also provide associated ages. 
The widely-used catalogue of \citet{Dias02} is a curated compilation of parameters from a large number of studies, which was obtained with a variety of methods and photometric systems. Another widely-cited study by \citet{Kharchenko13} presents an automated characterisation of the cluster population (known at the time), which was performed with all-sky 2MASS photometry \citep{Skrutskie06}. It represents a homogeneous set of parameters, but to a lesser precision than dedicated studies of individual objects.

The second data release of the European Space Agency (ESA) \textit{Gaia} mission \citep[DR2:][]{GDR2} represents the deepest all-sky astrometric and photometric survey ever conducted.
The \textit{Gaia} astrometry (proper motions and parallaxes) allows us to identify the members of clusters, and it has enabled the discovery of several hundreds of new objects. Combining parallaxes with the deep \textit{Gaia} photometry allows us to estimate cluster distances, ages, and extinctions on a large scale with unprecedented precision. Thus far, the largest study on this particular topic was conducted by \citet{Bossini19}, who derived parameters for 269 clusters (mostly nearby and well-populated). Despite the high precision of their results, this sample only constitutes less than 15\% of the clusters for which members can be identified with \textit{Gaia}.

The aim of the present work is to study the structure of the Galactic disc revealed by clusters of various ages. To this effect, we derived cluster parameters in a homogeneous and automatic fashion for $\sim$2000 Galactic clusters with members identified in the \textit{Gaia} data. 
In Sect.~\ref{sec:data} we present the input data and our list of reference clusters.
Section~\ref{sec:ann} describes the artificial neural network that we built and trained in order to estimate parameters. Section~\ref{sec:catalogue} introduces our catalogue of cluster parameters. In Sect.~\ref{sec:galaxy} we use this cluster sample to trace the structure of the Galactic disc. Section~\ref{sec:discussion} contains a discussion of the results, and Sect.~\ref{sec:conclusion} closes with concluding remarks.

\section{Data} \label{sec:data}

\subsection{Cluster members from \textit{Gaia} DR2} \label{sec:members}

We retained the probable members (probability >70\%) of 1481 clusters whose membership list was published by \citet{CantatGaudin20}, who estimated the membership probabilities for stars brighter than $G$=18 using the unsupervised classification scheme UPMASK \citep{KroneMartins14,CantatGaudin18tgas}. We collected the list of members provided by the authors for the recently discovered UBC clusters \citep{CastroGinard18,CastroGinard19,CastroGinard20}.

We also applied UPMASK to the 56 cluster candidates proposed by \citet{Liu19}. We were able to find secure members for 35 of them. These objects are listed in our catalogue as `LP', followed by the entry number given in \citet{Liu19}. 
Since UPMASK is not suited for very extended clusters, we took the list of members for the nearby clusters Melotte~25 (the Hyades) and Melotte~111 (Coma Berenices), which were derived from \textit{Gaia}~DR2 astrometry by \citet{GDR2hrd}.
In total, this compiled sample comprises $\sim$230,000 stars that are brighter than $G$=18, which belong to 2017 clusters.

\subsection{Reference clusters}  \label{sec:refOCs}

\begin{figure}[ht!]
\begin{center} \resizebox{0.5\textwidth}{!}{\includegraphics[scale=0.7]{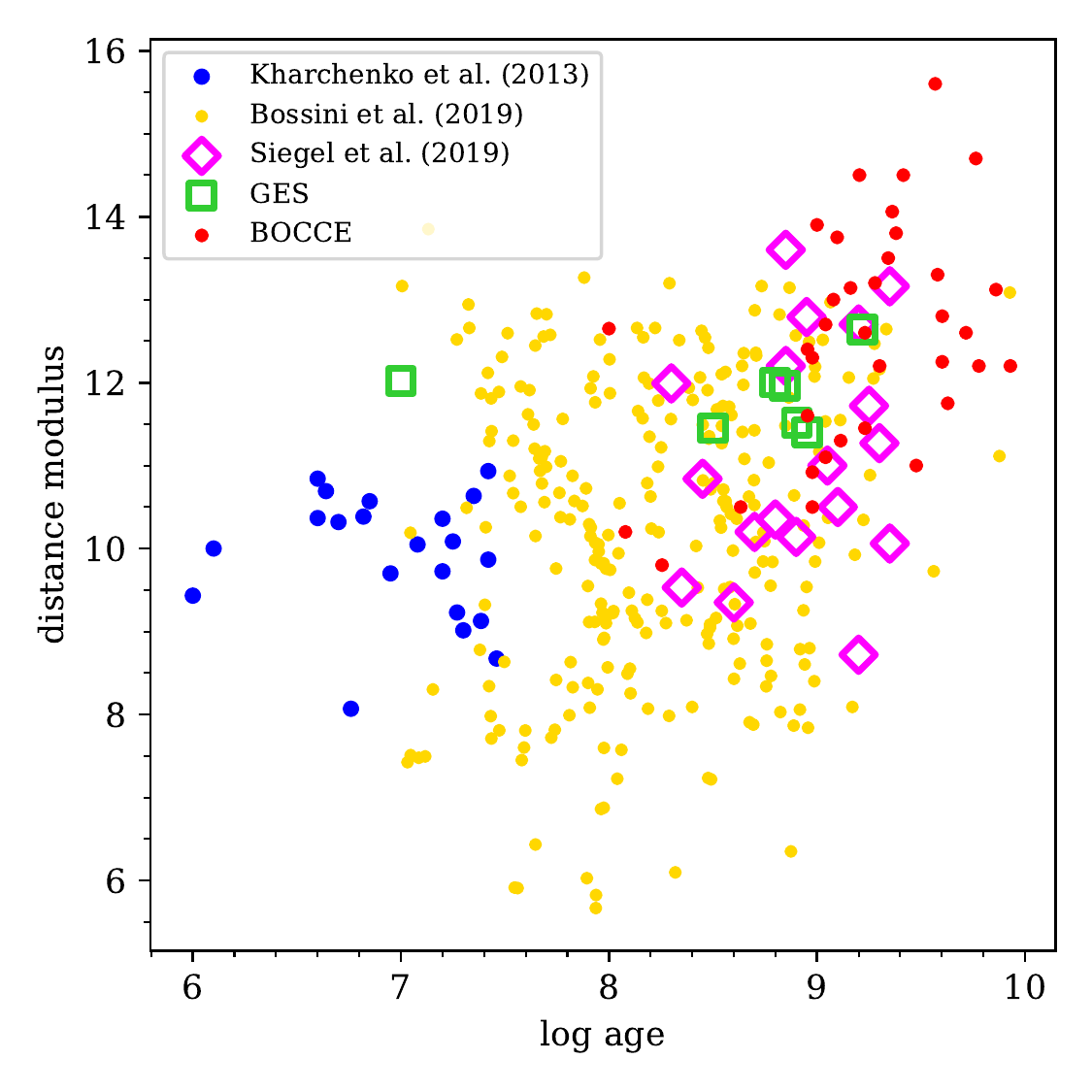}} \caption{\label{fig:refOCs} Age and distance modulus of our reference clusters (described in Sect.~\ref{sec:refOCs}).} \end{center}
\end{figure}

We compiled a list of 347 clusters with parameters (age, reddening, and distance modulus) that are known to a sufficient precision to be used as points of reference. Their ages and distances are shown in Fig.~\ref{fig:refOCs}. 
We strove to use a small number of reference studies to maximise homogeneity, while also covering the entire parameter space and privileging studies that employed \textit{Gaia} data for their membership selection.

The 269 clusters of \citet{Bossini19} represent the bulk of this reference set, and they constitute the largest homogeneous sample of cluster ages obtained from \textit{Gaia} data.
Their parameters were determined by fitting PARSEC isochrones \citep{Bressan12} with the Bayesian code BASE9 \citep{vonHippel06} to \textit{Gaia}~DR2 photometry of the cluster members identified in \citet{CantatGaudin18gdr2}.

This sample contains few clusters that are older than 1\,Gyr and few clusters that are more distant than 4\,kpc. We therefore supplemented the sample with 36 clusters from the BOCCE survey, which focuses mainly on old clusters, of which many are distant and characterised with a combination of deep photometry and high-resolution spectroscopy \citep{Bragaglia06,Bragaglia06be17,Tosi07,Andreuzzi11,Cignoni11,Donati12,Ahumada13,Donati14bocce,Donati15}.

Since these two samples contain very few clusters that are younger than $\log t \sim$7.5, we supplemented the training set with 21 young clusters with distances smaller than 1.5\,kpc and parameters that were taken from the catalogue of \citet{Kharchenko13}, which have visually well-defined colour-magnitude diagrams. We also included seven clusters that have been the subject of dedicated papers by the Gaia-ESO Survey:
NGC~3293          \citep{Delgado16}; 
NGC~4815          \citep{Friel14};   
NGC~6705          \citep{CantatGaudin14};    
NGC~6802          \citep{Tang17};   
Pismis~18         \citep{Hatzidimitriou19};  
Trumpler~20       \citep{Donati14}; and  
Trumpler~23       \citep{Overbeek17}.
We consider their parameters to be especially reliable due to the large number of radial velocities collected for these studies (allowing for good membership selections) and precise metallicities.

The \textit{Swift} UVOT Stars Survey provides cluster parameters for 49 clusters, which were studied with \textit{Gaia}~DR2 astrometry and isochrone fitting to near-ultraviolet photometry \citep[][]{Siegel19}. Eighteen of them are not present in the previously mentioned references, so we added them to our reference sample

\section{Cluster parameters and machine learning} \label{sec:ann}

Estimating the main parameters (age, distance modulus, extinction, and sometimes metallicity) of a star cluster is often done via isochrone fitting: A theoretical model of the sequence traced by a coeval group of stars in a two-dimensional colour-magnitude diagram (CMD) is compared to the observed distribution of stars.
Perhaps surprisingly, designing a robust and efficient automatic procedure for isochrone fitting is far from trivial. Observed CMDs of clusters do not simply follow a single sequence, but they feature unresolved binaries \citep[a problem addressed by the $\tau^2$ statistics of][]{Naylor06}, blue stragglers, broadened turnoffs \citep[][]{Marino18,Bastian18,Sun19,Li19,deJuanOvelar20}, and almost always contamination by field stars, which can also be taken into account with ad-hoc statistics \citep[as in e.g.][]{Monteiro10}. The stellar phases that provide the most clues about the age and distance of a cluster (its turnoff, red clump, and red giants) also happen to be the least populated parts of a CMD\footnote{The pre-main sequence of young clusters is also a good age indicator, but these low-mass stars are too faint to be observed in most objects.}, and they must be given a higher weight subjectively. 
The Bayesian code BASE9 \citep{vonHippel06,Jeffery16} relies on robust statistical principles and it allows for the use of prior knowledge (most importantly, a distance constraint provided by e.g. \textit{Gaia} parallaxes). However, its runs can be very time-consuming, it generally requires a large number of cluster members (it was in fact originally designed for globular clusters), and it is currently unable to deal with CMDs affected by differential extinction. The \texttt{ASteCa} package \citep{Perren15} uses a sophisticated approach with a modelling of a synthetic cluster from theoretical isochrones, but it is also relatively time-consuming and unable to deal with differential extinction and blue stragglers at present.

Isochrone fitting is therefore often performed by hand, which when done properly provides satisfactory results, but it is impractical to perform it on the samples of hundreds to thousands of clusters available from modern sky surveys. To address this problem and avoid direct comparisons with theoretical isochrones, we built a data-driven procedure to estimate the parameters of an unknown cluster based on its similarities with objects of known parameters. Although the age accuracy is ultimately tied to the reference values, which are derived from stellar evolution models, our approach has the advantage of putting all clusters on the same age scale and providing reliable relative ages.
Learning from labelled CMDs can be thought of as a generalisation of the empirically calibrated morphological age index, which allows for a quick estimate of a cluster age by measuring the magnitude difference between its turn off and red clump \citep[used by e.g.][]{Lynga82,Janes82,Janes88,Phelps94,Carraro94,Janes94,Friel95,Salaris04}, or the morphological age ratio \citep{AnthonyTwarog85,Twarog89}.

\subsection{Artificial neural network} \label{sec:anndef}

The increasing size and dimensionality of astronomical datasets have made machine learning increasingly popular in the field \citep[see e.g. the reviews by][]{Fluke19,Baron19}. 
Artificial neural networks (ANNs) are particularly popular due to their flexibility and performance at both classification \citep[e.g.][]{Ting18,CastroGinard18} and regression tasks \citep[e.g.][]{Leung19,Kounkel19,Boucaud20}.

\begin{figure*}[ht!]
\begin{center} \resizebox{1.\textwidth}{!}{\includegraphics[scale=0.6]{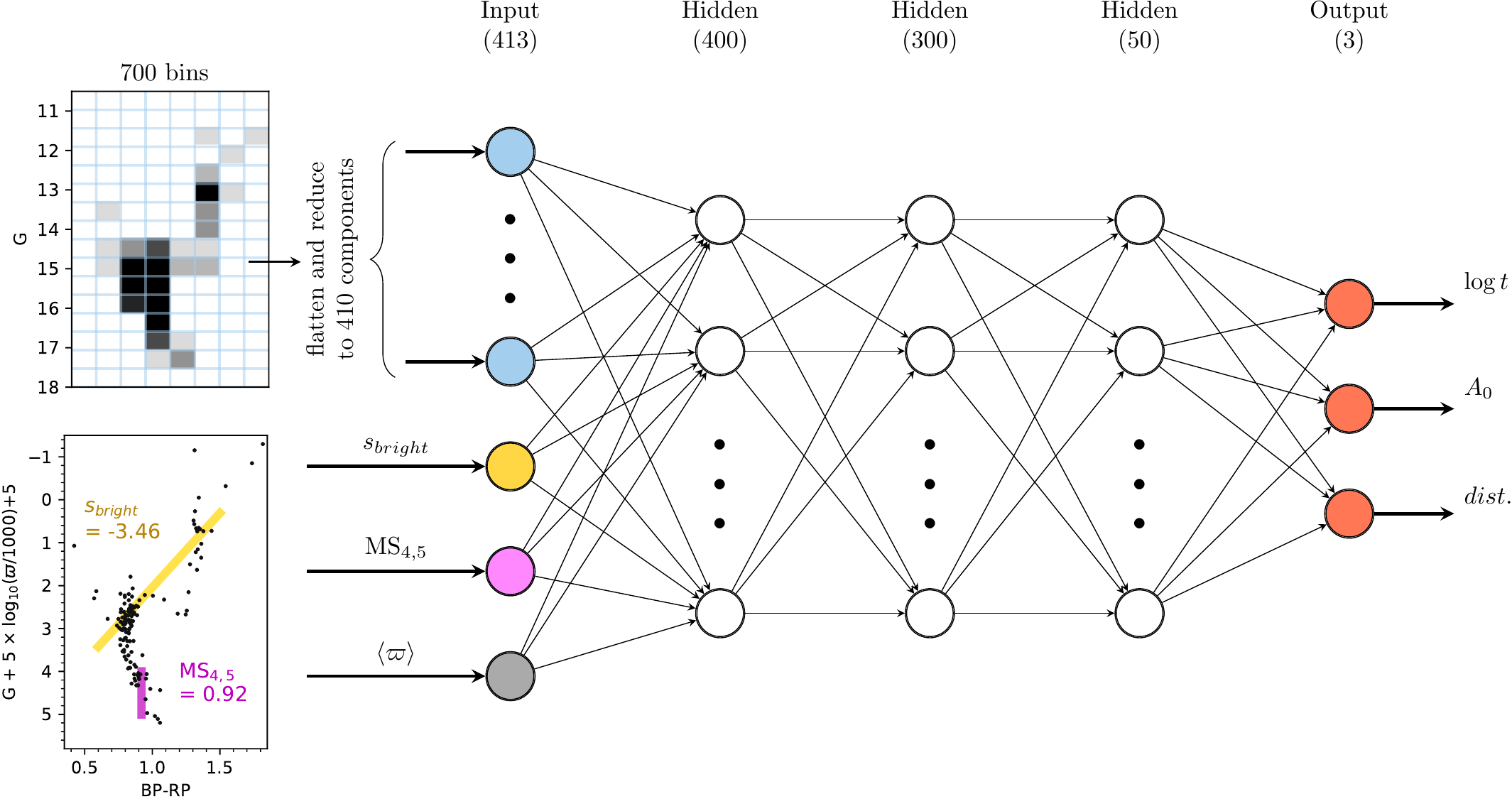}} \caption{\label{fig:finalNN} Architecture of our artificial neural network, indicating the width (number of nodes) of each layer. The example cluster is Haffner~22. The input quantities are described in Sect.~\ref{sec:anndef}.} \end{center}
\end{figure*}

An ANN is a system that maps the input observables to the target output quantities 
through a series of nodes. Here, the three targets are the cluster age, extinction, and distance modulus. Nodes are organised in layers, where every node receives input from the previous layer and output from a non-linear function of the input to the successive layer. For this work, we use a rectified linear unit (ReLU) . 
Formally, ANNs are universal approximators, which means that any continuous function can be approximated by an ANN with at least one hidden layer. Approximating a complex function might require a large number of nodes in the hidden layer, making the network slower to train and more prone to overfitting. An equivalent or better approximation can often be obtained with a smaller number of nodes if they are organised into several hidden layers in which each one contains an increasingly abstract representation of the data structure.
For this study we experimented with various architectures and settled on an ANN with three hidden layers, as is shown in Fig.~\ref{fig:finalNN}.

The main input observable that we provided to our ANN was a 2D histogram of the \textit{Gaia} colour-magnitude diagram of each cluster, with a bin width of 0.2\,mag in colour and 0.5\,mag in magnitude. The histogram was pre-processed before being fed to the ANN. We took the logarithm of the counts and scaled it so the most populated bin always had a value of 1. The entire histogram contains 700 bins. Applying a principal component analysis to the flattened histograms of our training set (described in Sect.~\ref{sec:trainingset}) shows that 99.9\% of the variance can be expressed with only 410 components. We therefore applied the transformation computed on the training set, which reduced the number of input quantities by nearly half, with a negligible loss of information.

We also provided the median parallax $\langle \varpi \rangle $ to the ANN, which is a strong predictor of distance, especially for the most nearby clusters. For each cluster, we provided two additional quantities estimated from the CMD (illustrated in Fig.~\ref{fig:finalNN}). The quantity $s_{bright}$ is the slope in the relation between colour and magnitude for the stars whose distance-corrected magnitude\footnote{The distance-corrected magnitude of a star is based on the cluster mean parallax $G=5\times\log_{10}(\langle \varpi \rangle/1000)+5$ and does not include correction for reddening.} is brighter than 4. This quantity strongly correlates with the cluster age. Finally, we denote MS$_{4,5}$ as the mean colour of stars whose distance-corrected magnitude is between 4 and 5. In this magnitude range, stars are always expected to be on the main sequence even in the oldest clusters, and their colour is a strong predictor of reddening.

If fewer than ten stars were available to estimate $s_{bright}$, we set it to an edge value of -10. If no stars were available for MS$_{4,5}$, which happens for distant and reddened clusters, we also set its value to -10. Thanks to their hidden layers, ANNs are able to approximate logical functions, which implicitly allows them to handle missing values.

\subsection{Training set} \label{sec:trainingset}

Our first attempts to estimate cluster parameters involved ANNs, which were trained with mock CMDs. Such systems were extremely good at recovering the input parameters of other mock CMDs, but overall they returned disappointing results when applied to real, observed, \textit{Gaia} CMDs. We therefore chose a data-driven approach that would not require us to generate mock clusters from theoretical models. Training machine learning procedures on labelled observed data is an increasingly common practice in various sub-fields of astronomy. For instance \citet{Ting18} trained an ANN to distinguish red giant branch stars from red clump stars, \citet{Leung19} determined elemental abundances with an ANN trained on high signal-to-noise ratio spectra, and \citet{Arnason20} identified new X-ray binary candidates in M31.

The basis of our training set are the clusters presented in Sect.~\ref{sec:refOCs}. A good training set must not only cover a wide range of parameters, but also be dense enough so that the ANN cannot memorise it and it must learn how the relevant features relate to the output. We performed data augmentation by creating variations of the reference clusters by artificially increasing their distance modulus and their extinction, by sub-sampling them, and by adding differential extinction.

The simulated distance modulus was randomly picked between 0.5\,mag smaller than the reference value and 16\,mag ($\sim$15.85\,kpc). 
We adjusted the simulated parallax accordingly and removed stars whose simulated $G$ magnitude was fainter than 18. 
To account for the uncertainties in the mean parallax, the local parallax zero-point variation, and to simulate the known zero-point offset in parallaxes \citep{Lindegren18,Arenou18}, we then subtracted 0.029\,mas and added a random offset that was uniformly picked between -0.05 and +0.05 mas. Adding noise to the simulated parallaxes is important so the ANN learns that for distant clusters, the distance modulus is mostly constrained by the CMD morphology and not by the parallax.

In order to cover a wide range in extinction, additional extinction was added up to $A_0$=5, using the polynomial relation presented by \citet{Danielski18} and \citet{GDR2hrd}. Differential reddening was added to half of the variations by first picking a random value between 0 and 1, setting the intensity of the differential reddening for this variation, then adding a random extinction picked between 0 and this maximum intensity.

Finally, we sub-sampled every reference cluster by picking a random number of stars, which went as low as ten, for every variation. In total we created 1500 versions of each reference cluster  and 3000 for the clusters with $\log t<7.4$ and $\log t>9.4$ since there are few of them in our reference sample.

Since the cluster members were selected based on their astrometry only, many clusters (especially the distant ones) include a fraction of field star contaminants. They were not removed from the training set, which means some training examples contain field contamination. 
The trained ANN is therefore able to deal with contamination in the non-reference clusters that we characterise in this study.

\subsection{Implementation and training}

We implemented the ANN on a desktop computer as a multi-layer perceptron regressor from the \texttt{scikit-learn} Python library \citep{scikit-learn}. The training was performed with the built-in ADAM solver \citep{Adam}. The \texttt{scikit-learn} implementation optimises the $R$-squared score defined as $R^2 = 1 - \frac{u}{v}$ where $u = \sum (y_{true}-y_{pred})^2$ is the residual sum of squares and $v = \sum ( y_{true}-\overline{y_{true}} )^2$ is the total sum of squares. A score of 1 would indicate a perfect prediction for all of the labels.

\begin{figure}[ht!]
\begin{center} \resizebox{0.5\textwidth}{!}{\includegraphics[scale=0.8]{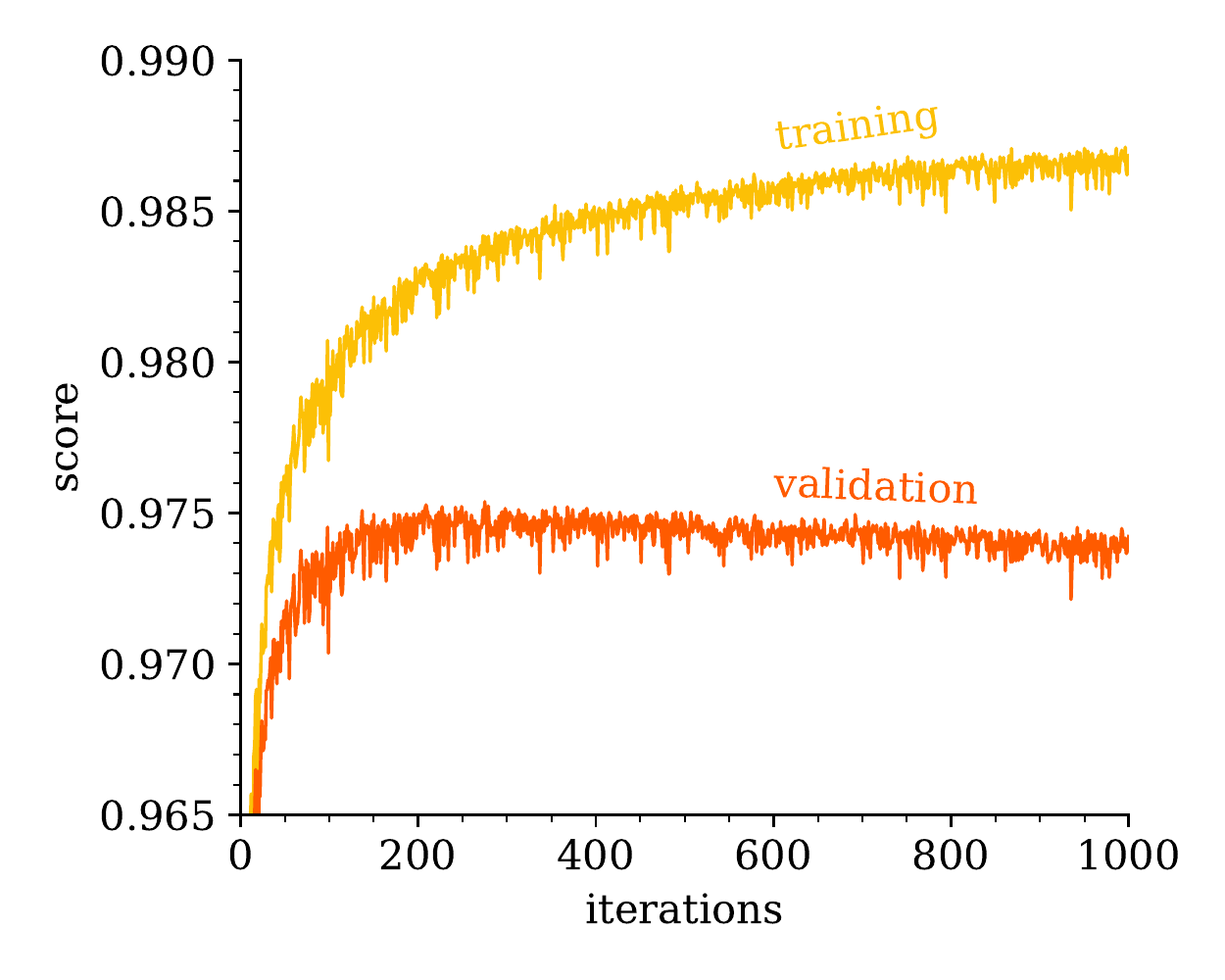}} \caption{\label{fig:score} Evolution of the training and validation scores with training iterations. The network used in this study is the result of 200 iterations.} \end{center}
\end{figure}

To make each training iteration faster and to alleviate the risk of the optimisation staying stuck in a local optimum, each iteration only used a random 20\% of the training set.
We built a validation set, which was created exactly like the training set, but containing other random variations of the reference clusters. We trained the ANN for 1000 iterations. 
At each iteration, we also verified the prediction of the ANN on the validation sample. We show in Fig.~\ref{fig:score} that although the training score steadily increases, the validation score reaches a maximum of around 200 iterations then it slowly decreases, which is a sign that the ANN starts overfitting. For the rest of this study, the ANN that we use is the one that was trained for 200 iterations.

\subsection{Performance on the validation set}

\begin{figure}[ht!]
\begin{center} \resizebox{0.45\textwidth}{!}{\includegraphics[scale=0.7]{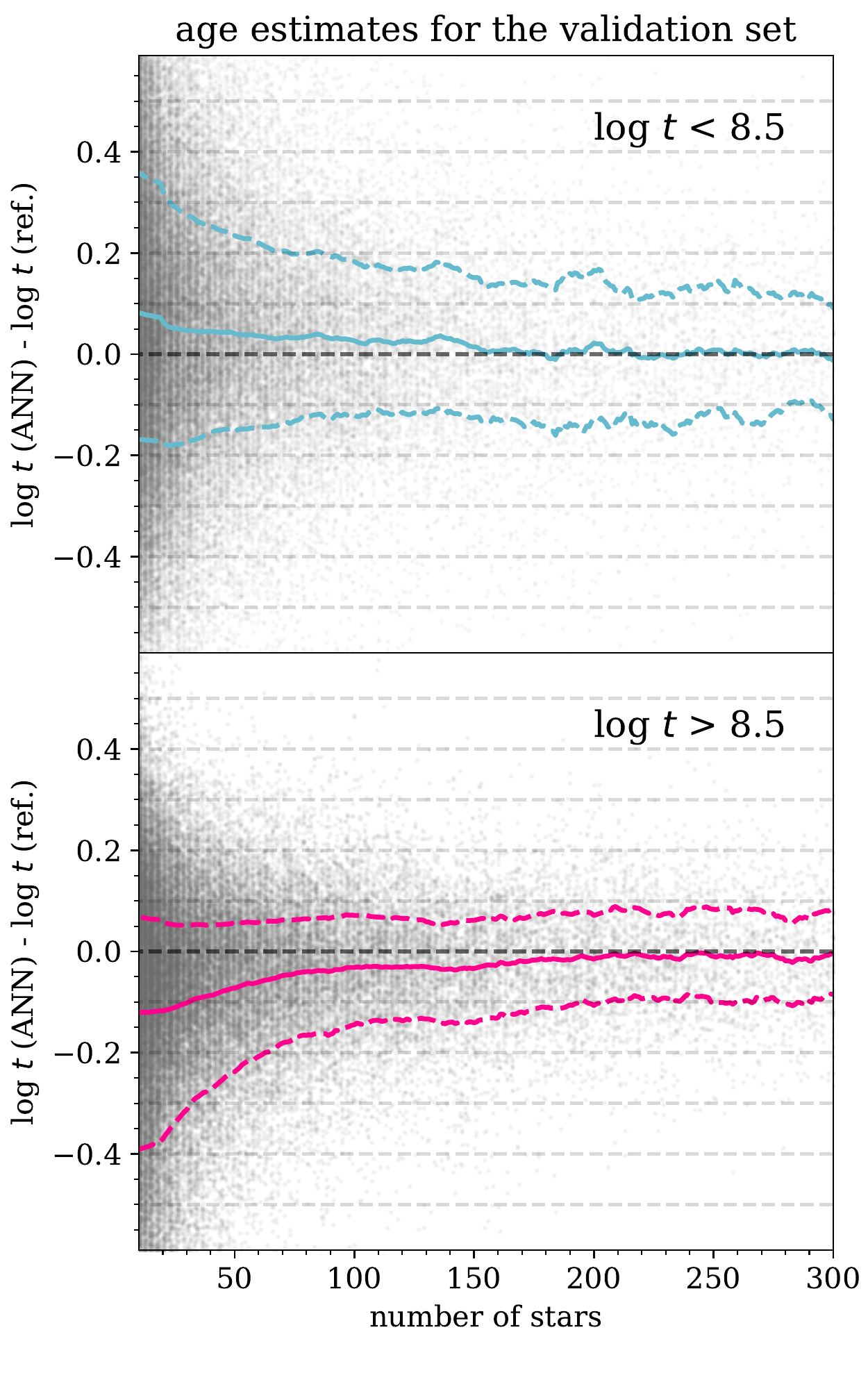}} \caption{\label{fig:agePREDerrors} Difference between the age estimate and the reference value for $\sim$120,000 validation samples, split in two age groups. The full line is a running mean. The dashed lines represent the upper and lower standard deviation. } \end{center}
\end{figure}

To assess the ability of the ANN to recover ages, extinctions, and distances, we investigated its performance on the validation set.
Figure~\ref{fig:agePREDerrors} shows the difference between the age estimated by the ANN and the reference value as a function of the number of stars. We see from this figure that young clusters with very few stars tend to have their ages slightly overestimated because the sparsely populated turn off appears fainter. Whereas for old clusters, the absence of red giants makes them appear younger. This is not specific to our machine learning approach, but rather a general limitation of using CMDs to estimate cluster ages. In practice, less than 10\% of our observed clusters have fewer than 20 members. In a successive step (Sect.~\ref{sec:catalogue}), we also flag the clusters whose CMDs are too sparse and/or too blurry to show a meaningful pattern.

Overall, the uncertainty on the determination of $\log t$ ranges from 0.15 to 0.25 for young clusters and from 0.1 to 0.2 for old clusters.
For the extinction and distance modulus, the precision of the ANN also depends on the number of stars, but only marginally on the age of the cluster. The typical uncertainty of $A_0$ ranges from 0.1 to 0.2\,mag, and the typical distance modulus uncertainty ranges from 0.1 to 0.2 ($\sim$5\% to 10\% distance uncertainty).

If we assume that the reference values represent the ground truth, then these mean differences indicate the precision of our procedure. However the scatter encompasses both the uncertainties due to our methodology and the uncertainties of the reference parameters.

At the beginning of training, the weights of the ANN are initialised to random values. Every training run therefore converges to a slightly different final state. We have verified that the difference between several networks trained for 200 iterations with the same training set is negligible.

\section{The catalogue of cluster parameters}  \label{sec:catalogue}

We applied the trained ANN to estimate the parameters of all 2017 clusters mentioned in Sect.~\ref{sec:members}. We visually inspected the CMD of every cluster, with theoretical isochrones corresponding to the estimated parameters. For the large majority of them, the result looked satisfactory and closely matched the result that would have been obtained by a human expert. In 61 cases, the parameters had to be adjusted manually in order to better match the aspect of the CMD with a PARSEC isochrone \citep{Bressan12} of solar metallicity. The reason why the ANN performed poorly on these objects is not clear -- they do not correspond to a specific age or distance range -- and might be due to field contaminants. 
The parameters proposed by the ANN were still close enough to make this manual correction faster than having to pick an isochrone without a suggested starting point. We flagged these 61 objects in our catalogue.

We also flagged 81 clusters whose CMD is too blurred and reddened. They mostly distribute close to the Galactic plane in the direction of the Galactic centre, and most of them are known embedded clusters. Some of these objects include NGC~1579, which is associated with the Northern Trifid HII region, or the young massive clusters Westerlund~1 and Westerlund~2.  

We further flagged 69 objects for which the CMD is too sparse to estimate meaningful parameters from photometry. 
Finally, we used literature values for three objects with a clear enough CMD but where the ANN failed to recover good parameters. Two of them are the very nearby Hyades (Melotte~25) and Coma Ber (Melotte~111), whose distance modulus is out of the range covered by our training set. We set their parameters to the values quoted by \citet{GDR2hrd}. The third cluster is Gaia~2 for which our only members are red giant branch stars. We took its parameters from \citet{Koposov17}.

We end up with 1867 clusters with reliable parameters.
We provide the list of all investigated clusters with their mean parameters and corresponding flags as an electronic table.

\subsection{Comparisons with the literature}

\begin{figure*}[ht!]
\begin{center} \resizebox{0.99\textwidth}{!}{\includegraphics[scale=1.]{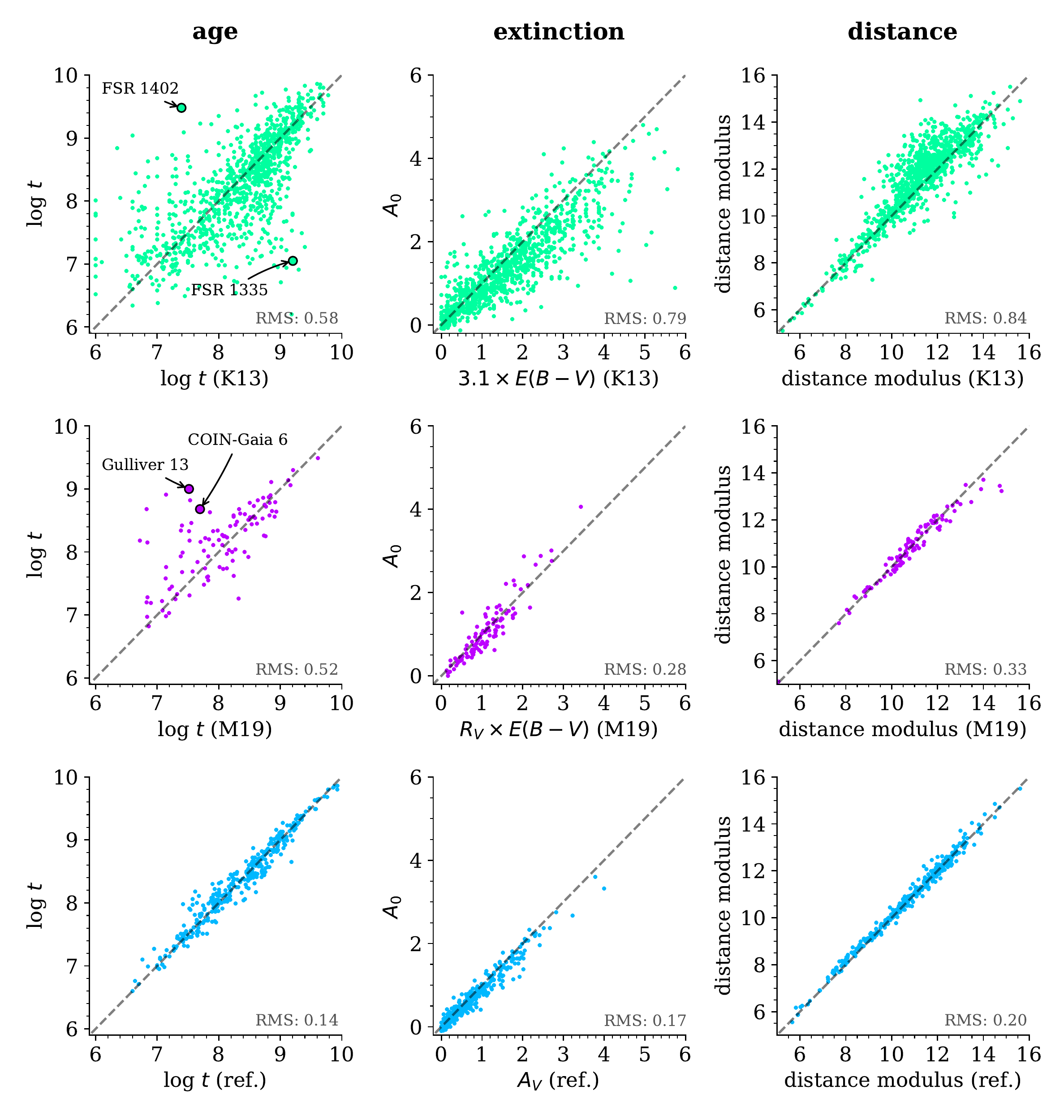}} \caption{\label{fig:comparisons_with_lit} Top row: Comparison of the parameters for the clusters in common with \citet{Kharchenko13}. Middle row: Comparison of the parameters for the clusters in common with \citet{Monteiro19}. The CMDs and isochrones for the labelled clusters are shown in Fig.~\ref{fig:fourNewFits}. Bottom row: Comparison between our ANN parameters and the literature references presented in Sect.~\ref{sec:refOCs}. All panels display the root mean square (RMS) difference.} \end{center}
\end{figure*}

\begin{figure}[ht!]
\begin{center} \resizebox{0.5\textwidth}{!}{\includegraphics[scale=1.]{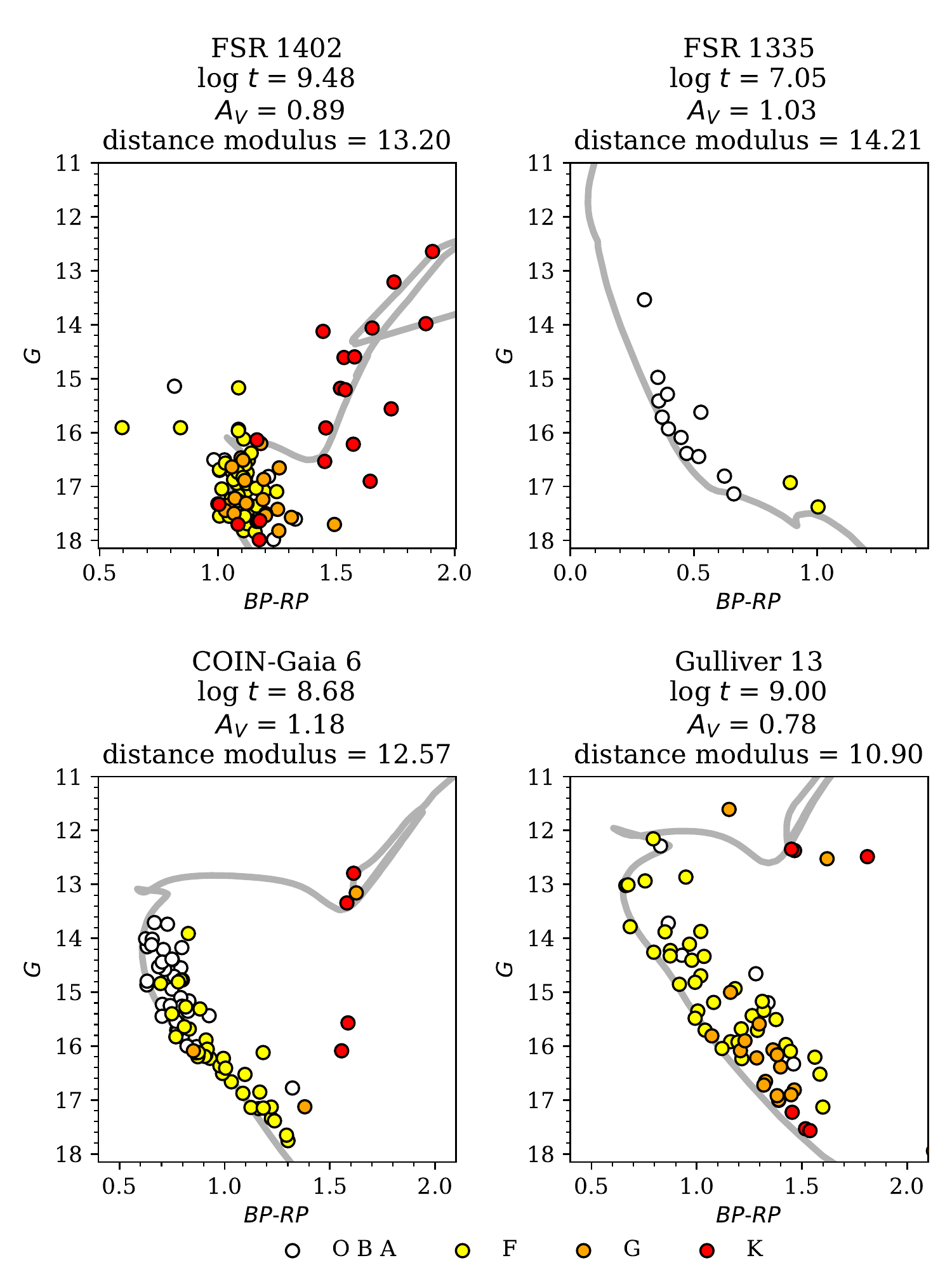}} \caption{\label{fig:fourNewFits} Colour-magnitude diagram, colour-coded by spectral type from the effective temperatures of \texttt{StarHorse} \citep{Queiroz18,Anders19} for the four clusters labelled in Fig.~\ref{fig:comparisons_with_lit}. The lines are PARSEC isochrones of solar metallicity.} \end{center}
\end{figure}

In the top row of Fig.~\ref{fig:comparisons_with_lit}, we show comparisons between our recovered parameters and the values listed by \citet{Kharchenko13} (hereafter K13), which were obtained by isochrone fitting to 2MASS photometry \citep{Skrutskie06}. 
Many of the clusters for which K13 lists old ages while we find young ages are very reddened objects, where the bright turnoff stars have been mistaken by K13 for a red branch (e.g. FSR~1335, whose CMD is shown in Fig.~\ref{fig:fourNewFits}). Conversely, the cleaner membership and the distance constraint provided by the \textit{Gaia} astrometry show that objects such as FSR~1402 (also shown in Fig.~\ref{fig:fourNewFits}) are evolved clusters. Since FSR~1335 is young, sparse, and distant, any estimate of its age from just \textit{Gaia} photometry of its brightest members is affected by large uncertainties. It is, however, evident that it is not an old cluster. 
Our procedure generally returns lower extinctions than K13. This could be due to our choice of defining $A_0$ as the extinction corresponding to the blue edge of the sequence in a CMD, before the effect of differential reddening, rather than determining the value for which the isochrone passes through the middle of the sequence.

A comparison with the parameters recently published by \citet{Monteiro19} is shown in the middle panel of Fig.~\ref{fig:comparisons_with_lit}. The authors relied on \textit{Gaia}~DR2 to select cluster members and constrain their distance, thus explaining the better agreement to our results. Several clusters still have discrepant age estimates, almost all of them are due to the presence of red stars that we consider to be cluster members. Two of them are labelled in Fig.~\ref{fig:comparisons_with_lit}, and their CMDs are shown in Fig.~\ref{fig:fourNewFits}. 

The bottom row of Fig.~\ref{fig:comparisons_with_lit} shows comparisons with the reference values for the clusters we used to build the training set (presented in Sect.~\ref{sec:refOCs}). The fact that we do not exactly recover the reference parameters is a good sign because it shows that the ANN did build an approximation of the relation between observables and cluster parameters, rather than memorising the aspect of reference clusters. The largest age discrepancies affect a handful of clusters for which \citet{Bossini19} list ages $\log t \sim 7.6$, while our ANN estimates $\log t \sim 7.9$. These objects are too distant for their pre-main sequence stars to be visible, so the main age constraint is the ill-defined location of their turn off.

\subsection{Composite Hertzprung-Russell diagram}

\begin{figure*}[ht!]
\begin{center} \resizebox{0.95\textwidth}{!}{\includegraphics[scale=1.]{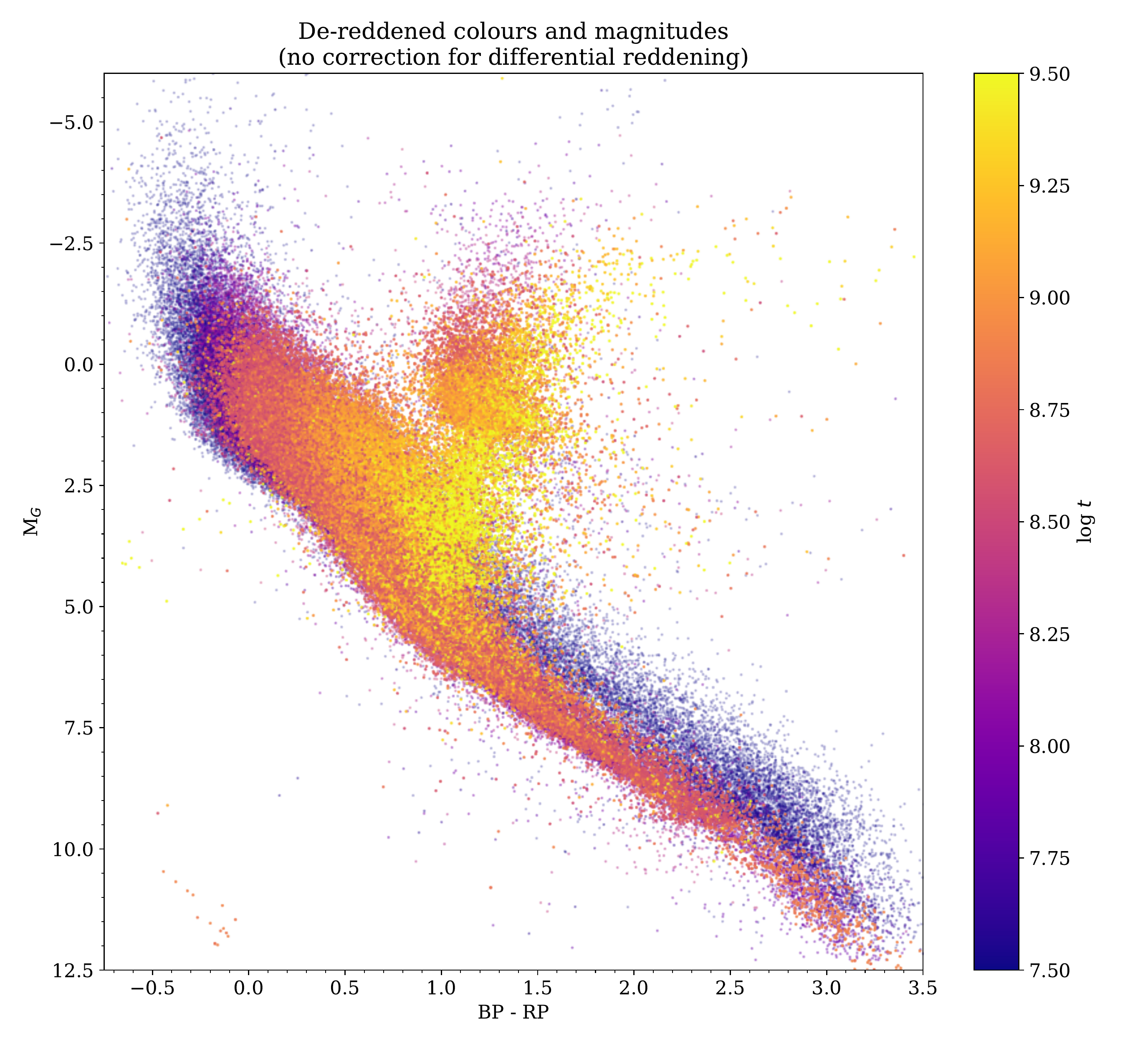}} \caption{\label{fig:HRD} Comprehensive Hertzprung-Russell diagram including 1867 clusters, colour-coded by cluster age.} \end{center}
\end{figure*}

Having an estimate of $A_0$ for each cluster, we corrected the observed colours and magnitudes for interstellar extinction by inverting the relations given in \citet{Danielski18} and \citet{GDR2hrd}. We then corrected $G$ for distance modulus. The comprehensive Hertzprung-Russell diagram (HRD), which is made up of 1867 clusters, is shown in Fig.~\ref{fig:HRD}.

Since a single value of extinction was used for each cluster, this HRD is still affected by differential extinction, which is especially apparent in the elongation of the red clump. A few white dwarfs can be seen. They belong to the very nearby Hyades (Melotte~25), Coma Ber (Melotte~111), and Praesepe (NGC~2632). In the lower right part of the diagram, the presence of pre-main sequence stars is clearly visible in clusters younger than $\log t$$\sim$8.

All of the cluster members used in this study have an apparent $G$ magnitude that is brighter than 18. Since most of the old and very populated clusters are distant objects (e.g. Berkeley~32 or Collinder~261), few old stars with M$_G$ $>$ 5 are visible in the HRD.

\subsection{Limitations and potential improvements}

Although age, distance, and extinction are the parameters that contribute most to the aspect of a cluster in a CMD, metallicity also plays a role, especially for the coolest stars. Some studies leave it as a free parameter when performing isochrone fitting, but it is common to keep it fixed to an assumed value, as a wrong value mostly affects the reddening and only has a small impact on ages\footnote{The morphological age index of \citet{Salaris04} includes a $\log t$ correction of 0.07 per dex of metallicity.}. 
In this study we did not train the ANN to estimate metallicities, but the training set spans a large range in metallicity. Given that we fed the ANN a coarsely binned representation of the CMD, and given the strong degeneracy between metallicity and extinction, it is unlikely that our ANN could be used to make meaningful estimations of this quantity. An additional issue is that only a relatively small fraction of clusters have homogeneous and precise abundance determinations from high-resolution spectroscopy, which are and often from inhomogeneous sources \citep[a problem discussed by][]{Heiter14}, meaning that such a machine learning procedure would have to rely on a training set built with mock data.

Since we binned the millimag-precision \textit{Gaia}~DR2 photometry \citep{Evans18} into a grid with a resolution of  0.2\,mag in colour and 0.5\,mag in $G$ magnitude, our approach is obviously not able to take advantage of the finest features observed in some \textit{Gaia} CMDs. For the best-defined clusters, isochrone fitting procedures \citep[e.g.][]{Naylor06,vonHippel06,Monteiro10} are able to extract more information from the CMDs. We experimented with a finer binning of the CMD, but the size of the training set and the exponential increase in training time made this impractical. In the future, procedures employing an adaptive kernel density estimation might help to overcome this issue.

The use of ground-based photometry, especially at non-optical wavelengths, and value-added catalogues containing astrophysical parameters for individual stars \citep[][]{Andrae18,Anders19} could help to provide better constraints on the cluster parameters.
Colour-magnitude diagrams are not an optimal approach for young clusters, especially when their pre-main sequence stars are not visible and the only age constraint is the colour of the bluest, most massive, identified member. They can also be affected by significant inhomogeneous extinction or feature small age spreads. When spectroscopic measurements are available, the lithium depletion boundary method (LDB) can provide a better constraint than photometry \citep[e.g.][]{Barrado04,Jeffries05}, but it can return older ages than CMD fitting \citep[e.g. 21\,Myr versus 7.5\,Myr in][]{Jeffries17}.
\citet{Lyra06} have reported and discussed systematical differences between the nuclear ages, for main sequence stars, and contraction ages for pre-main sequence stars. \citet{Randich18} performed a homogeneous analysis of seven clusters younger than $\sim$100\,Myr, making use of three different sets stellar evolution models of ($J$,$H$,$K_s$,$V$) photometry and LDB models. They confirm that much of the scatter found in the literature for the age of these objects can be attributed to the use of different models or the choice of photometric passbands included in the isochrone fitting.
An additionnal issue affecting young and embedded clusters is that star-forming regions are sometimes known to present anomalous reddening laws that differ from the general interstellar medium \citep[e.g.][]{Feinstein73,Vazquez96,Hur12,Kumar14}, while the present study employs the same fixed reddening law for all clusters. However, \citet{Jordi10} remark that varying the extinction law within the range reported by \citet{Fitzpatrick07} has a negligible effect on the \textit{Gaia} photometry.

Another promising approach to deriving cluster ages is the analysis of stellar rotation \citep[so-called gyrochronology,][]{Barnes07}, which presents the advantage of allowing age estimates for main sequence stars, and up to several billions of years \citep[e.g.][]{Meibom15,Douglas19}.
A spectacular application of this method is the characterisation of the recently discovered Pisces-Eridanus stream \citep[][]{Meingast19}. While it had been previously claimed (based on a single red giant with an uncertain membership status) that the structure could be 1\,Gyr old, \citet{Curtis19} show that 154 main sequence stars with available rotation periods exhibit a similar rotation pattern to the Pleiades ($\sim$120\,Myr). \citet{Curtis19} also point out that although theoretical models have so far been unable to perfectly fit the observed loss of stellar angular momentum with age, empirical comparisons with benchmark clusters of a known age can already provide robust constraints. The Transiting Exoplanet Survey Satellite \citep[TESS,][]{Ricker15} provides an all-sky survey from which light curves can be obtained, and many of its targets are cluster members \citep{Bouma19}. In our sample, several clusters\footnote{The `Class C' clusters UBC~605, 610, 625, 632, 642, and 649 from \citet{CastroGinard20} are compact in astrometric space but their CMDs are sparse and blurry.} are located at high Galactic latitudes and only contain late-type stars, but their ill-defined turnoff and the absence of red clump stars make it impossible to constrain their age. The increase in available training data (from e.g. TESS) and the flexibility of machine learning procedures, allowing for missing values and the empirical combination of measurements of a different nature, will make it possible to constrain the ages of such difficult objects.

\section{Galactic structure}  \label{sec:galaxy}

\begin{figure*}%
    \centering
    \subfloat{\includegraphics[scale=0.7]{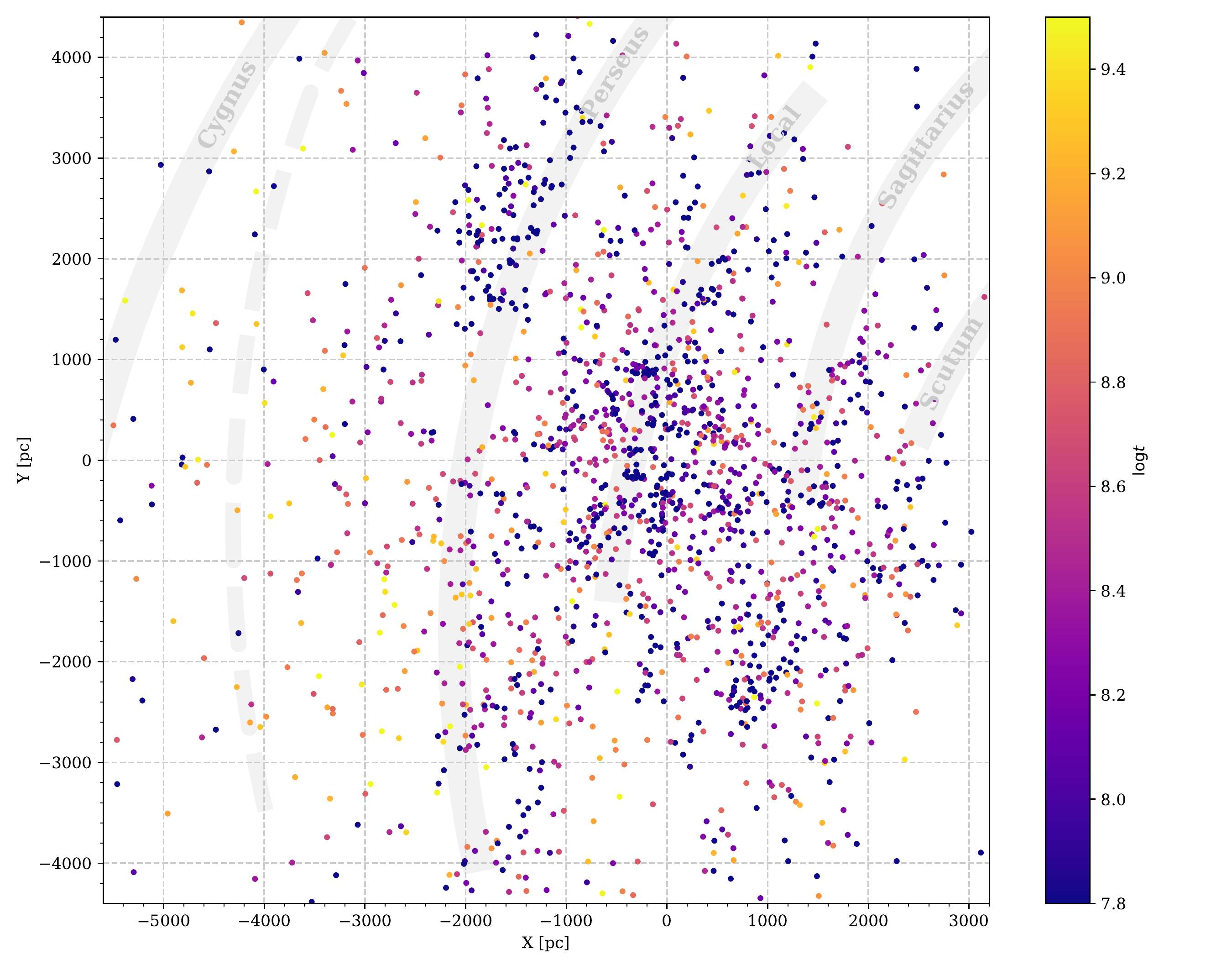}}%
    
    \subfloat{\includegraphics[scale=0.6]{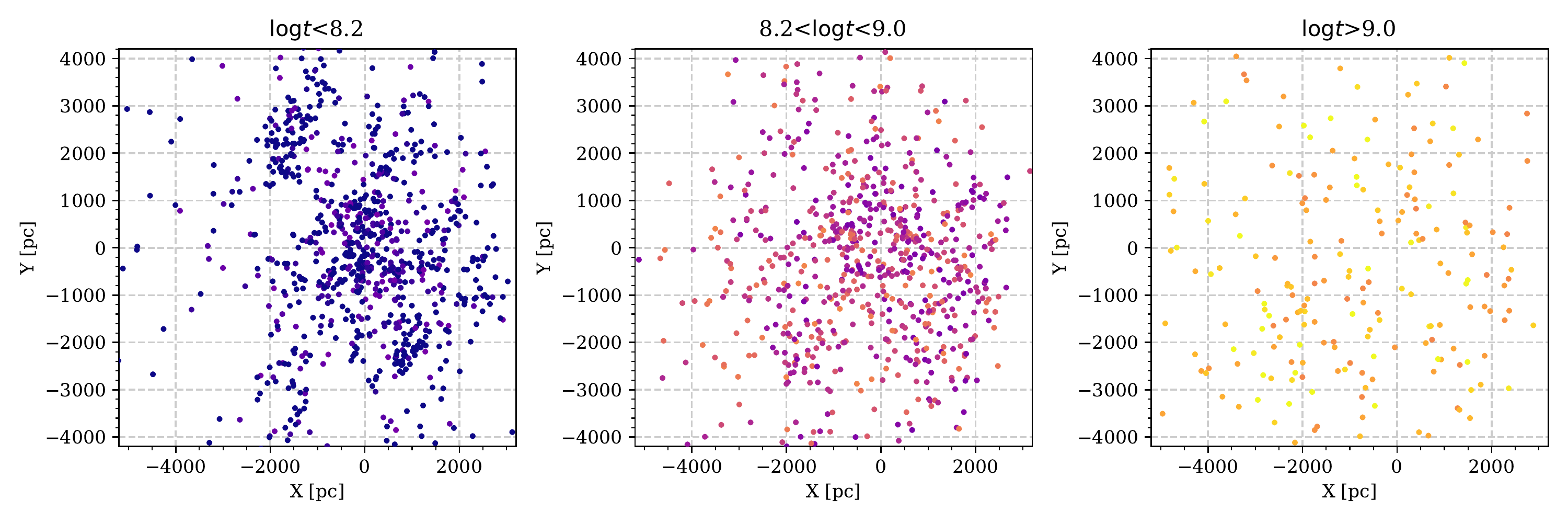}}%

    \caption{\label{fig:XY} Projection on the Galactic plane of the locations of clusters with derived parameters, colour-coded by age. The top panel shows all ages. The shaded area shows the spiral arm model of \citet{Reid14}. The dashed arm is the revised path of the Cygnus arm in \citet{Reid19}. The bottom row splits the sample into three age groups. The Sun is at (0,0) and the Galactic centre is to the right. The most distant objects were left out of the plot.}%
\end{figure*}

\begin{figure*}[ht!]
\begin{center} \includegraphics[scale=0.7]{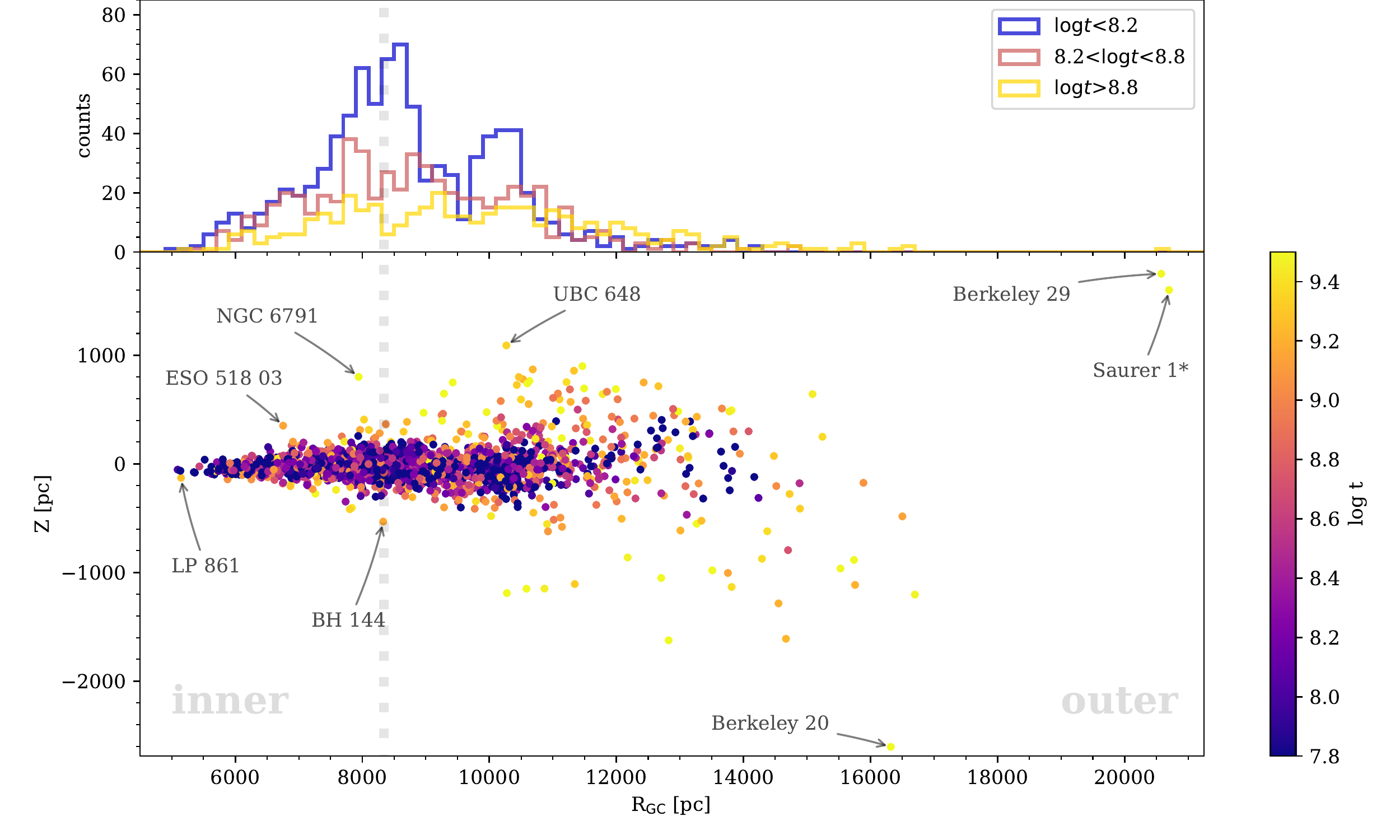} \caption{\label{fig:RgcZ} Top: Galactocentric distribution for three age groups. Bottom: Distance from the Galactic plane against Galactocentric distance, colour-coded by age, for the clusters with derived parameters. The vertical dotted line shows the assumed Solar value of R$_{\mathrm{GC}}$=8340\,pc \citep{Reid14}. Our catalogue lacks Saurer~1 members, so we took its parameters from \citet{Carraro03}. } \end{center}
\end{figure*}

Using the derived distance modulus, we computed the (X, Y, Z) cartesian coordinates\footnote{The Sun is at the origin. We note that X increases towards the Galactic centre, Y is in the direction of Galactic rotation, and Z is in the direction of the Galactic north pole.} of all clusters with available parameters. We show the projection of the cluster distribution on the Galactic plane in Fig.~\ref{fig:XY}. We also computed the Galactocentric radius R$_{\mathrm{GC}}$, assuming a Solar Galactocentric distance of 8340\,pc\footnote{The most precise and recent estimate \citep{Gravity19} proposes a slightly smaller radius of $\sim$8180\,pc.}, which is the value adopted by the spiral arm model of \citet{Reid14}. The R$_{\mathrm{GC}}$ versus Z distribution is shown in Fig.~\ref{fig:RgcZ}.

\begin{figure*}[ht!]
\begin{center} \includegraphics[scale=0.75]{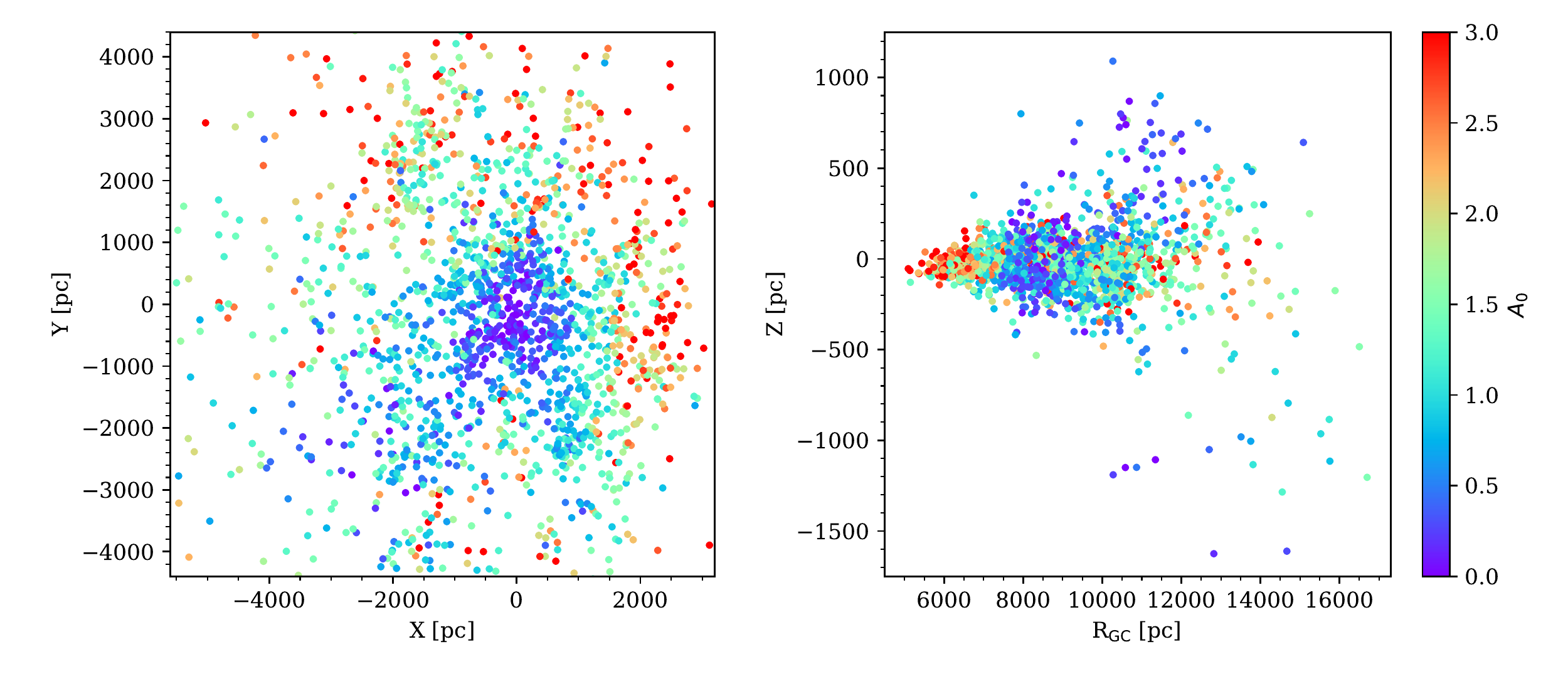} \caption{\label{fig:a0} Distribution of clusters in Galactic XY coordinates (left) and altitude versus Galactocentric radius (right), colour-coded by extinction $A_0$. In both panels, a few distant outliers were left out of the plotting window.} \end{center}
\end{figure*}

We show the distribution of extinction in Fig.~\ref{fig:a0}. The sample of known clusters reaches much larger distances in the direction of the outer disc, especially for objects located far above the plane, but it is still limited by interstellar reddening at low Galactic latitudes.

\subsection{Spiral structure}

The spatial distribution of young clusters is known to correlate with the location of the spiral arms in the Milky Way \citep{Morgan53,Becker70,Dias05}. The projection of the cluster distribution is shown in Fig.~\ref{fig:XY}. Its general aspect is similar to Figure~11 in \citet{CantatGaudin18gdr2}, where groups of young clusters distribute preferentially along the locations of spiral arms delineated by \citet{Reid14}, but with important gaps and discontinuities.

In the region covered by the present study, the updated spiral arm model of \citet{Reid19} is virtually identical. Most differences affect the first Galactic quadrant at distances that our sample of clusters does not reach, with the notable exception of the outer, Cygnus, arm. For this arm, \citet{Reid19} fitted a significantly different location with a pitch angle of 3$^{\circ}$ and R$_{\mathrm{GC}}$$\sim$11 to 13\,kpc in the anticentre direction \citep[compared to 13.8$^{\circ}$ and 13 to 15\,kpc in][]{Reid14}. We show the revised arm as a dashed line in Fig.~\ref{fig:XY}.

Our sample of \textit{Gaia}-confirmed clusters only contains very few objects with R$_{\mathrm{GC}}$ > 12\,kpc. 
The top panel of Fig.~\ref{fig:RgcZ} exhibits two clearly visible peaks in the young cluster distribution, corresponding to the local arm and the Perseus arm. The Cygnus arm is not visible due to the lack of available tracers. \citet{Camargo15} estimated the distance to several embedded clusters that were identified in \textit{WISE} infrared images \citep{Wright10}, and they propose that they trace the Cygnus arm at a Galactocentric distance of 13.5 to 15.5\,kpc, which agrees with the more distant \citep[][]{Reid14} model.

It has been noted \citep[e.g.][]{Vazquez08} that the Perseus arm traced by clusters appears to be interrupted in the Galactic longitude range of $\sim$140$^{\circ}$ to 160$^{\circ}$. 
Many clusters have been discovered in the Perseus arm region in the past decade, including two dedicated searches in \textit{Gaia}~DR2 \citep[][]{CantatGaudin19coin,CastroGinard19}, but all of them were found around the gap, rather than inside of it. This region of low density is visible around (X,Y)=(-2000\,pc,+1000\,pc) in the maps displayed in Fig.~\ref{fig:XY}.

\begin{figure*}[ht!]
\begin{center} \includegraphics[scale=0.6]{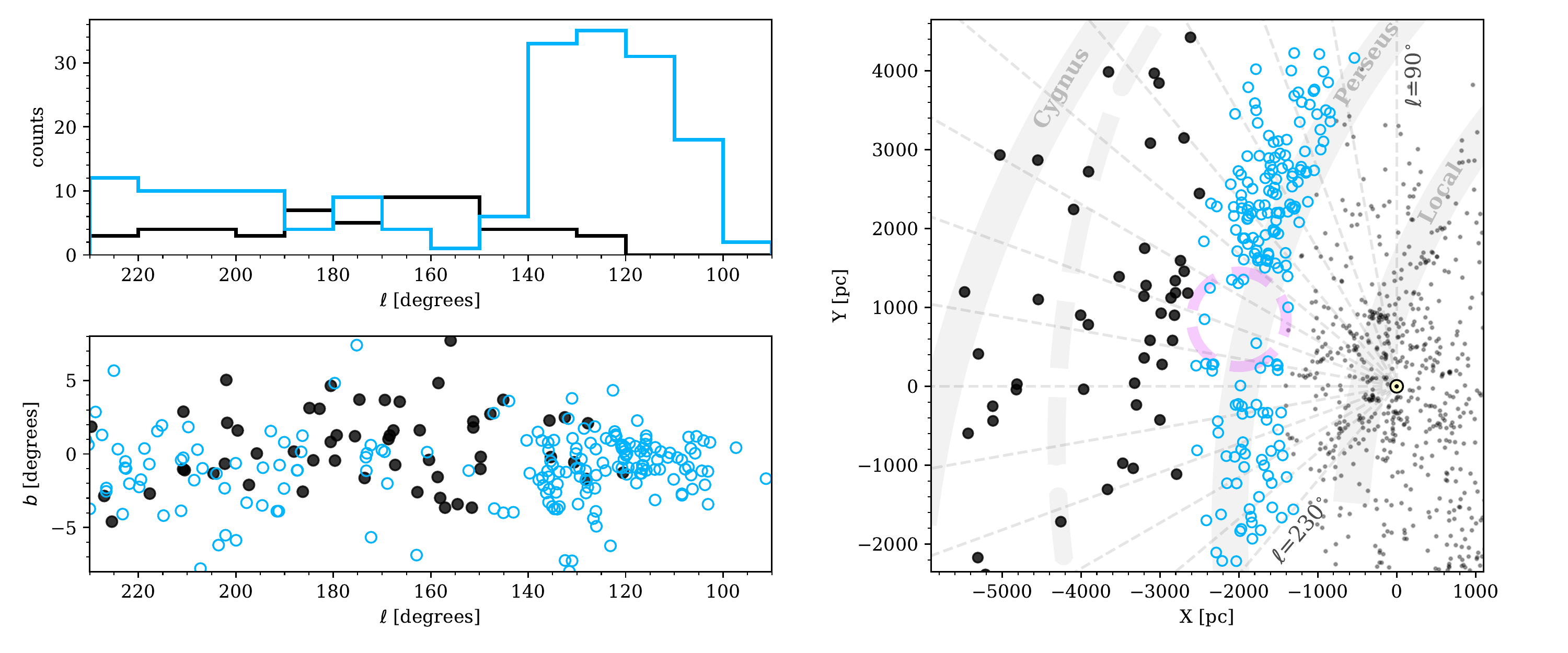} \caption{\label{fig:perseusGap} Top left: Distribution of Perseus arm clusters (arbitrarily selected as 10\,kpc < R$_{\mathrm{GC}}$ < 11\,kpc) in cyan, and more distant clusters in black. Bottom left: Galactic coordinates of the same clusters. Right: Locations of the same clusters projected on the Galactic plane. The dashed circle has a diameter of 1200\,pc. } \end{center}
\end{figure*}

A natural explanation for the lack of detected objects in this direction could be that our view is obscured by interstellar dust, but this range of Galactic longitude does not correspond to a known region of high extinction \citep[e.g.][]{Lallement19}. The strongest argument against extinction being responsible for this gap is illustrated in Fig.~\ref{fig:perseusGap}. While the number of clusters located at the distance of the Perseus arm drops for $\ell$$\sim$140$^{\circ}$ to 160$^{\circ}$, the number of known clusters located behind the arm increases. 
It can be seen in Fig.~\ref{fig:a0} that the clusters located beyond the gap are only moderately reddened, with values of $A_0$$\sim$1.5\,mag.
This in fact suggests that the Perseus gap is a window of relatively lower extinction.

This gap is visible in the distribution of other young tracers, which are traditionally associated with spiral arms, and is in fact present in the HII map of \citet{Becker70} as well as in the HI map of \citet{Spicker86}, although the authors do not comment on it. The distribution of HII regions used by \citet{Hou14} to trace the spiral structure is interrupted in the same region, and the gap can be seen (tentatively) in the Cepheid distribution of \citet{Skowron19} as well as in the OB stars shown by \citet{Romero19}, \citet{Poggio19}, and \citet{Jardine19}, and the high-mass star-forming regions of \citet{Reid14} and \citet{Reid19}. 

A possibly similar gap, which is not as clear however, can be observed in the Sagittarius arm (Fig.\ref{fig:XY}), with an under-density of young clusters around (X,Y)=(+1000\,pc,-1000\,pc).
Studying clusters in kinematical space could indicate that these arms are fragmenting, which is a phenomenon routinely seen in $N$-body simulations \citep[e.g.][]{RocaFabregas13,Grand14,Hunt15}, and this would show that the Milky Way is not a grand design spiral galaxy, but rather a flocculent one. 

We also see that the interarm region between the local arm and the Perseus arm is not as clear in the third quadrant as in the second quadrant, which is in agreement with \citet{Moitinho06} and \citet{Vazquez08}, who propose that the local arm extends towards the Perseus arm along the $\ell$=245$^{\circ}$ line.
The presence of young clusters in the region between the Perseus and outer arms can also be interpreted as the trace of interam spurs, as reported by \citet{MolinaLera19} and suggested by the HII maps of \citet{Hou14}. Such features are visible in external spiral galaxies \citep[e.g.][]{Corder08,Elmegreen18} and naturally occur in numerical simulations \citep[e.g.][]{Shetty06,Dobbs06,Pettitt16}.


\subsection{Scale height} \label{sec:scaleheight}
The fact that old clusters tend to be found at higher Galactic altitudes (further away from the plane) than young clusters has been noted by numerous observers \citep{vandenBergh58,vandenBergh80,Janes88,Janes94,Phelps94,Friel95}, and this is visually obvious from our Fig.~\ref{fig:RgcZ}. 
The main cause for the thickening of the Galactic disc is the gradual velocity scatter, which is introduced by gravitational interactions with giant molecular clouds \citep[first theorised by][]{Spitzer51,Spitzer53}, although it is now understood that the effects of the spiral structure, Galactic bar, warp, and even minor mergers have contributed to the vertical heating of the disc \citep[see e.g. the recent study of][and references therein]{Mackereth19}.

Various analytical parametrisations of the vertical density distribution are used in the literature \citep{vanderKruit88,Dobbie20}. A simple form often used for the cluster distribution is the exponential profile:

\begin{equation}
    N(Z) = \frac{1}{h_z} \exp{ \left( - \frac{|Z-\langle z \rangle|}{h_z}  \right) }
,\end{equation}

\noindent where $\langle z \rangle$ is the mean offset of the Galactic plane with respect to the Sun and the $h_z$ parameter is called the scale height. Many authors perform a fitting of the scale height in bins of age or Galactocentric radius. Rather than binning, we modelled it with a power-law dependence on age ($t$) and a linear dependence on the Galactocentric radius:

\begin{equation} \label{eq:zh}
    h_z = k + a \times \left( \frac{t}{100\,\mathrm{Myr}} \right)^{\alpha} + \rho \times (\mathrm{R}_{\mathrm{GC}} - \mathrm{R}_{\mathrm{GC},\odot}) .
\end{equation}

\begin{figure*}[ht!]
\begin{center} \includegraphics[scale=0.6]{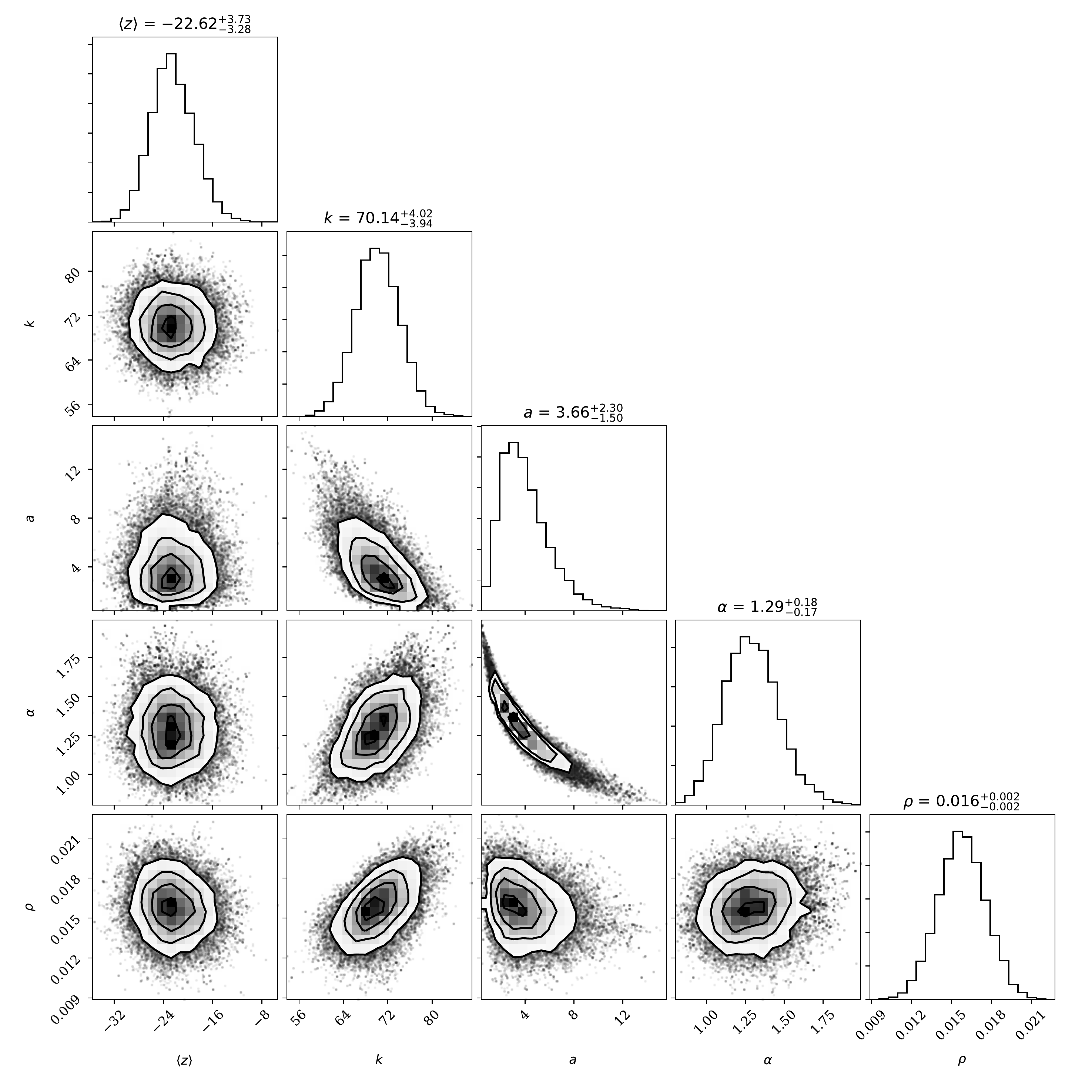} \caption{\label{fig:emceeCornerPlot} Markov chain Monte Carlo sampling of the posterior distribution for the scale height model presented in Sect.~\ref{sec:scaleheight}, showing the last 2000 iterations of 32 walkers (64,000 points). } \end{center}
\end{figure*}

We sampled the parameter space using the Markov chain Monte Carlo sampler \texttt{emcee} \citep{ForemanMackey13}, with flat priors on all parameters. The resulting posterior distribution is shown in Fig.~\ref{fig:emceeCornerPlot}.

The first free parameter in our model is $\langle z \rangle$, that is, the mean altitude of the entire sample considering that the Sun sits at altitude 0. The best fit value is $\langle z \rangle$=$-23\pm3$\,pc, corresponding to a solar displacement of $z_0$=$23\pm3$\,pc. This value is in line with estimates from star counts from
\citet{Juric08} ($25\pm5$\,pc),
\citet{Chen99} ($28\pm6$\,pc),
\citet{Chen01} ($27\pm4$\,pc), or
\citet{MaizApellaniz01} ($24\pm2$\,pc), for instance. We remark that studies making use of young tracers tend to report a slightly smaller solar displacement, which can be seen in \citet{Karim17} ($17\pm2$\,pc) or \citet{Reed06} ($19.6\pm2.1$\,pc), for instance, and previous estimates based on clusters such as in \citet{Buckner14} ($18.5\pm1.2$\,pc) and \citet{Joshi07} (13 to 20\,pc) reported smaller values. The altitude of the Galactic mid-plane is known to vary with Galactocentric radius \citep[sometimes called corrugation, see e.g.][]{Gum60,Lockman77,Spicker86,CantatGaudin18gdr2}, which might be an additional reason why different samples yield slightly different values\footnote{We refer the interested reader to \citet{Karim17}, who compiled a list of over 60 estimates published since 1918.}.

In the Solar neighbourhood, where the typical cluster age is $\sim$100\,Myr, the cluster scale height of the best-fit model is 74$\pm5$\,pc, which is marginally
compatible with the 64$\pm$2\,pc of \citet{Joshi16}.
Our best-fit value of $\rho = 0.016\pm0.003$ (18\,pc per kpc) is in good agreement with the value of 0.02 reported by \citet{Buckner14}.

We also find that the scale height increases to several hundreds of parsecs for old clusters \citep[also reported by][]{Janes94,Froebrich10,Buckner14}, with a power-law index of $\alpha = 1.3\pm0.2$. The mechanism often invoked to explain the steeper increase at higher ages is that clusters whose orbits do not reach high $Z$ are destroyed at higher rates, which is due to crossing paths more often with giant molecular clouds \citep{Moitinho10,Buckner14}. 
\citet{Friel95} remarked that some old clusters reach such high altitudes that the encounter responsible for perturbing their orbit would likely disrupt the cluster in the process. Although \citet{Gustafsson16} have shown that some clusters might survive such strong perturbations, there are no quantitative arguments to support that this mechanism is the only reason for the increase in scale height.

The phenomenon of heating has been studied more thoroughly for field stars than for clusters, but almost all studies have been performed in velocity space rather than positional space, making direct comparisons difficult. The time dependence of the vertical velocity dispersion in the Solar neighbourhood is often modelled as a power law ($\sigma_v \propto t^{\alpha}$). Theoretical models predict values of $\alpha<0.3$ \citep{Hanninen02}, while observations of field stars suggest an age exponent of $\alpha \sim 0.5$ \citep[e.g.][]{Wielen77,Holmberg09,Aumer16,Sharma20}, showing that other mechanisms have contributed to vertical heating such as mergers \citep{Martig14} or more efficient scattering by giant molecular clouds in the young Milky Way \citep{Ting19}. We refer the interested reader to Sect.~5.3.2 of \citet{BlandHawthorn16}, who discuss recent estimates of the age-velocity dispersion relation.

The age-scale height relation we derive in this study cannot be directly compared to the age-velocity relation. It is not clear how a power-law increase of index 0.5 in velocity dispersion translates in positional space. The details of the relation between maximum velocity and maximum excursion from the Galactic plane ($Z_{max}$) depend on the assumed Galactic potential. For the \texttt{MWPotential2014} which was shipped with \texttt{galpy} \citep{Bovy15}, the relation is close to $Z_{max} \propto v ^{1.3}$, implying a steeper time dependence than a power law of index 0.5.

Radial migration and heating can also cause clusters to reach higher altitudes: Due to the shallower potential of the outer disc, their vertical velocity allows particles to reach larger excursion from the plane when their guiding radius is shifted outwards. If inward-migrating clusters are destroyed at higher rates than outward-migrating clusters \citep[as suggested by e.g.][]{Anders17}, then the mean Galactocentric radius and mean altitude of surviving clusters is expected to increase with age.
Radial heating also contributes because particles on elliptical orbits reach higher altitudes near their apocentre.

The scale height of very young clusters appears to be rather large in the outer disc, with several of our clusters younger than 200\,Myr reaching altitudes of 300\,pc. Although the distances of these distant objects are less precise than for more nearby clusters, these results are compatible with the infrared findings of \citet{Camargo15}, who report seven embedded, and therefore very young, clusters that are further than 500\,pc from the Galactic plane at R$_{\mathrm{GC}}$$\sim$14\,kpc. Our simple model assumes a linear increase in the scale height with Galactocentric radius, but \citet{Kalberla07} and \citet{Kalberla08}, who could trace atomic hydrogen out to much larger distances than our cluster sample, show that the flaring of HI gas outside the Solar circle is better reproduced with an exponential function, and \citet{Wang18} used a quadratic function.

Finally, if cluster disruption rates are lower in the outer disc, one would also expect scattering rates to be lower. Mathematically, this could be modelled by modifying equation (\ref{eq:zh}) to allow the index $\alpha$ to vary with $\mathrm{R}_{\mathrm{GC}}$. Including radial migration, heating, and disruption rates varying with $\mathrm{R}_{\mathrm{GC}}$ and $Z$ would make the model overly complicated and poorly constrained, with highly degenerate parameters. 

Characterising the velocity distribution of clusters is out of the scope of this paper, but it would provide further insight on the processes of migration, heating, and disruption. Detailed chemical studies through high-resolution spectroscopy can also shed light on the origin of clusters. The old, metal-rich object NGC~6791 is a well-known case of a cluster migrating from the inner disc \citep{Jilkova12,Carraro14,MartinezMedina18}, but lesser-known or newly discovered clusters with discrepant altitudes (such as BH~144 or UBC~648, labelled in Fig.~\ref{fig:RgcZ}) might also shown evidence for radial migration.

\subsection{Galactic warp}

The Galactic mid-plane is known to deviate from the geometrical $b=0^{\circ}$ plane in the outer disc, which is a phenomenon called warp.
The warping of the Galactic plane is particularly visible in the HI gas distribution \citep{Burke57,Kerr57,Westerhout57,Levine06,Kalberla07} and is now known to be a common feature in disc galaxies \citep[e.g.][]{Sancisi76,Briggs90,SanchezSaavedra03}. 
The warp is also visible in the distribution of molecular clouds \citep{Wouterloot90}, dust \citep{Marshall06}, stars \citep{LopezCorredoira02,Moitinho06,Vazquez08,Reyle09,Amores17,Chrobakova20}, and stellar kinematics \citep{Poggio18,Schonrich18}; additionally, it was recently investigated by tracing the distribution of classical Cepheids \citep{Skowron19,Chen19}. These young \citep[$\sim$20 to 120\,Myr:][]{Efremov78,Bono05,Senchyna15} and bright stars are visible at large distances and allow for precise distance determinations.

\begin{figure*}[ht!]
\begin{center} \includegraphics[scale=0.75]{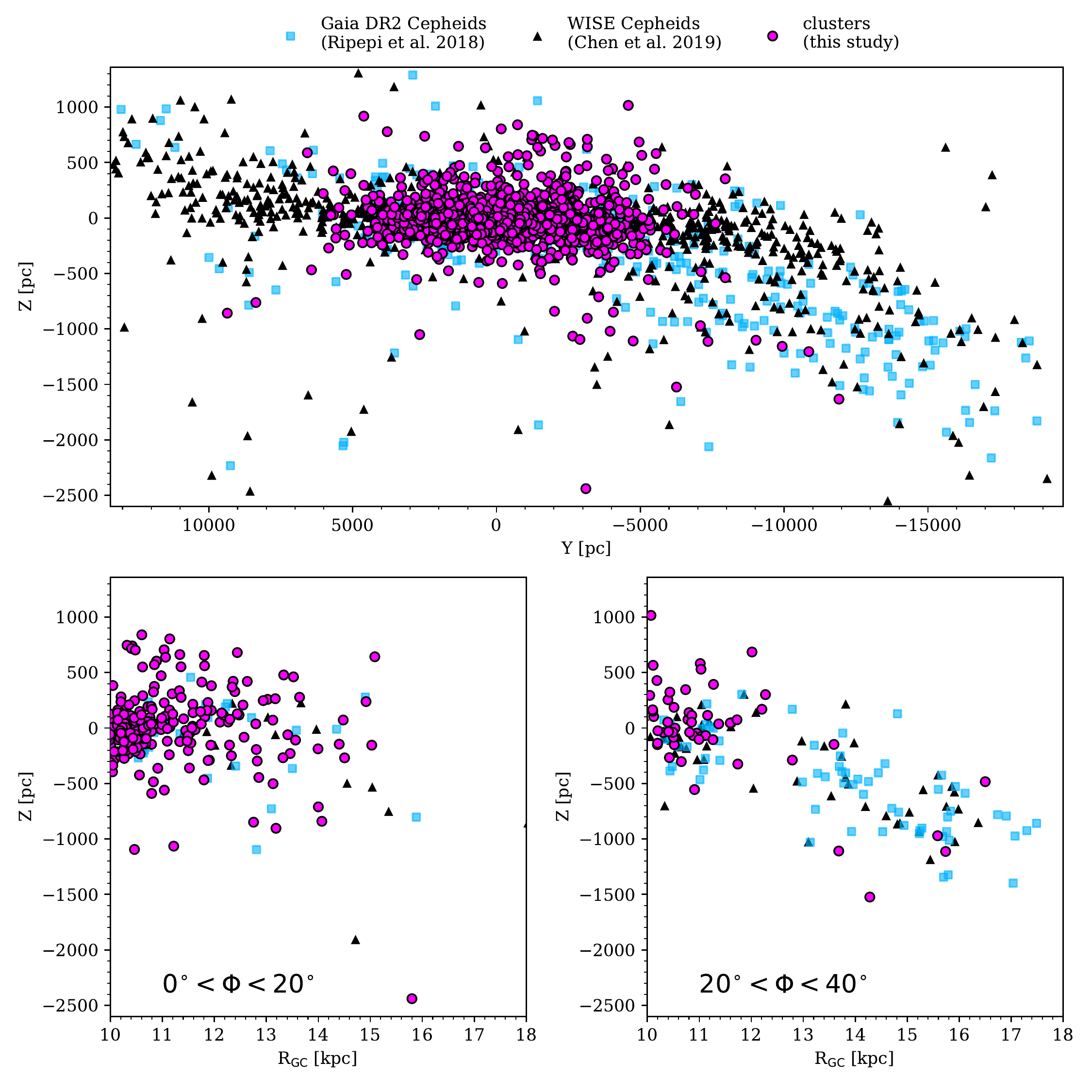} \caption{\label{fig:warp} Top: Y versus Z coordinates of our cluster sample and the Cepheids from \citet{Ripepi19} and \citet{Chen19}. Bottom: Galactocentric distance versus altitude Z in two ranges of Galactocentric angular coordinates, both in the third Galactic quadrant. } \end{center}
\end{figure*}

In Fig.\ref{fig:warp} we compare the location of known clusters with classical Cepheids. The lower panels only include tracers in two bins of Galactocentric angular coordinates $\Phi$, where $\Phi$=$0^{\circ}$ is the line passing through the Galactic centre and the Solar location, and $\Phi$ increases in the opposite direction to Galactic rotation \citep[convention used in e.g.][]{Ripepi19,Skowron19}.
The distant clusters in the third Galactic quadrant are on average older than 1\,Gyr, and they follow the same southward trend as the young Cepheids. The number of known distant clusters is unfortunately too small to allow us to verify whether the Cepheid warp and the old cluster warp still coincide for $\Phi>40^{\circ}$. In particular, no known clusters are located in the region of the northern warp.

\section{Discussion} \label{sec:discussion}

Among the clusters for which we can derive parameters, the closest to the Galactic centre is Ruprecht~126 ($\log t$=8.11; R$_{\mathrm{GC}}$=5230\,pc). 
Several known clusters might be located even deeper in the disc, according to their small parallax and apparent location, but their CMDs are too sparse and blurry to allow us to derive meaningful parameters and to constrain their distance with photometry. The deepest known clusters would be BH~222 \citep[also studied by][]{Piatti02} and Gulliver~41, both of which are at R$_{\mathrm{GC}}$<3\,kpc and lack estimated parameters in our catalogue.

We label in Fig.~\ref{fig:RgcZ} several old clusters that stand out as outliers. One of them is the well-studied NGC~6791, an old metal-rich cluster whose likely origin is the bulge or the inner disc. Berkeley~20, Berkeley~29, and Saurer~1 are also well-known distant objects, which are currently located far from the Galactic plane. The object UBC~648 is a recently discovered cluster \citep{CastroGinard20}, and it is also located far from the Galactic plane.

The cluster LP~861 was only recently discovered \citep{Liu19} and is one of the innermost old clusters known. Other intermediate-age or old clusters were recently identified in the \textit{Gaia}~DR2 data, such as UBC~307, UBC~310, UBC~339, LP~866, and UFMG~2, which are all located at R$_{\mathrm{GC}}$<6.5\,kpc and at very low altitudes. 
The only such objects known before \textit{Gaia} were NGC~6005 \citep{Piatti98}, NGC~6583 \citep{Carraro05}, Ruprecht~134 \citep{Carraro06}, and Teutsch~84 \citep{Kronberger06}. 
These objects deserve further investigation in order to understand how they can survive to reach old ages in such a dense environment. They might be on very elliptical orbits, have recently migrated inwards, or their initial mass may have been sufficient for them to remain gravitationally bound.

We cannot presently probe the structure of the outer disc (e.g. the trace of the Cygnus arm or the geometry of the warp) with the sample of clusters identified in \textit{Gaia} (with $G$<18). As is visible in Fig~\ref{fig:RgcZ}, very few clusters are known at R$_{\mathrm{GC}}$>14\,kpc and no clusters are known beyond 16.5\,kpc, with the exception of Berkeley~29 and Saurer~1 which are near R$_{\mathrm{GC}}$$\sim$20\,kpc. This lack of available tracers is due, at least in part, to an obscured line of sight preventing us from identifying distant objects near the Galactic plane. A near-infrared \textit{Gaia}-like mission \citep{Hobbs16,Hobbs19} would allow us to see through dust clouds and reveal obscured structures and embedded clusters. 
The upcoming ground-based LSST \citep{LSST09,Ivezic19} will reach stars seven magnitudes fainter than \textit{Gaia}, and it is expected to provide proper motions better than 1\,mas\,yr$^{-1}$ down to $G$$\sim$24, allowing one to push the boundaries of cluster detection further than presently possible.

We note however that the distant outer disc clusters, especially in the third quadrant, are not strongly affected by extinction (Fig.~\ref{fig:a0}).\ This suggests that the drop in density is not just an observational bias, but also a sign that few clusters populate the distant outer disc. Stellar population studies typically locate the disc truncation radius near 14\,kpc \citep{Robin92}, 15\,kpc \citep{Ruphy96}, or 16\,kpc \citep{Amores17}.
Due to the uncertainty on the completeness of our sample in the outer disc, we did not attempt to fit a radial density profile or try to identify a cut-off Galactocentric radius, but the observed cluster distribution visually agrees with a cut-off point near 14\,kpc.
The objects Berkeley~29 and Saurer~1, which are on the far edge of the disc, would therefore be outliers on very perturbed orbits, rather than representants of a cluster population forming at extreme Galactocentric distances. 
On the other hand, several distant disc clusters were recently discovered with a combination of \textit{Gaia} data and deep ground-based photometry by authors searching for satellite systems \citep{Koposov17,Torrealba19}. The lack of clusters beyond R$_{\mathrm{GC}}$$\sim$16\,kpc could therefore be an observational bias that future studies will be able to fill in.

This study focuses on the present-day location of clusters. The \textit{Gaia}~DR2 catalogue also allows us to determine proper motions for all of them and, therefore, estimate tangential velocities.  \citet{Soubiran18} have obtained mean radial velocities for several hundreds of clusters using the \textit{Gaia} Radial Velocity Spectrometer \citep{Cropper18} and shown a smooth increase in vertical velocity dispersion with age. Further insight can be gathered by supplementing the scarce \textit{Gaia} radial velocities with observations from other surveys \citep[e.g.][with APOGEE and GALAH data]{Carrera19}. 
Although \textit{Gaia}~DR3 will contain significantly more radial velocities than DR2 \citep{Brown19}, the \textit{Gaia} spacecraft only has limited spectroscopic capabilities.
Ground-based spectroscopic surveys such as APOGEE \citep{Majewski17}, \textit{Gaia}-ESO \citep{Gilmore12,Randich13}, GALAH \citep{deSilva15}, LAMOST \citep{Cui12}, or the upcoming WEAVE \citep{Dalton12} and 4MOST \citep{deJong12,Guiglion19} will provide additional observations allowing for the full characterisation of the 3D velocities of many more objects, and they will shed light on the dynamical processes that drive the evolution of the spiral structure and the heating of the Galactic disc.

\section{Summary and conclusion} \label{sec:conclusion}

This study relies almost exclusively on \textit{Gaia}~DR2 data. We characterise clusters whose members were identified with \textit{Gaia} astrometry. We use an artificial neural network to estimate the age, distance modulus, and interstellar extinction of each cluster from the \textit{Gaia} photometry of its members and their mean \textit{Gaia} parallax. The training set was built using observed clusters with reliable parameters.

After visually inspecting the colour-magnitude diagrams and verifying the consistency of the parameter estimates, we end up with 1867 clusters with reliable parameters. The 3D distribution of clusters traces the structure of the Galactic disc, with warping and flaring in the outer disc. We clearly observe the known increase in cluster scale height with age. Various mechanisms contribute to this increase, and the current cluster locations are not sufficient at disentangling the effects of heating, migration, and location-dependent disruption rates. Establishing the 3D velocity vector and characterising the orbital parameters of clusters and their dependence with age will provide further insight on the evolutionary history of the Milky Way.

Projected on the Galactic plane, the locations of young clusters roughly align along the expected spiral pattern, and especially the local and Perseus arms. We argue that the apparent interruption in the Perseus arm is physical, and it is not due to an observational bias introduced by interstellar extinction. More kinematical data is needed in order to determine whether the Perseus arm is in the process of fragmenting. 
Our present sample does not contain a sufficient number of distant clusters to trace the path of the outer arm or constrain the geometry of the warp in the outer disc.

The catalogue presented in this paper is the largest homogeneous analysis of cluster parameters performed with \textit{Gaia} data so far, with almost two thousand objects. 
The continuous discovery of new clusters and the development of data-driven methods that are capable of including other photometric passbands, astrophysical parameters from value-added catalogues, or rotation periods will allow for more precise and accurate cluster parameter estimates as well as a consistent account of observational errors.

\section*{Acknowledgements}

We thank the referee for useful suggestions that helped clarify this paper.
This work has made use of data from the European Space Agency (ESA) mission \textit{Gaia} (www.cosmos.esa.int/gaia), processed by the \textit{Gaia} Data Processing and Analysis Consortium (DPAC, www.cosmos.esa.int/web/gaia/dpac/consortium). Funding for the DPAC has been provided by national institutions, in particular the institutions participating in the \textit{Gaia} Multilateral Agreement. 
This work was supported by the MINECO (Spanish Ministry of Economy) through grant ESP2016-80079-C2-1-R and RTI2018-095076-B-C21 (MINECO/FEDER, UE), and MDM-2014-0369 of ICCUB (Unidad de Excelencia 'María de Maeztu'). TCG acknowledges support from Juan de la Cierva - Formaci\'on 2015 grant, MINECO (FEDER/UE). FA is grateful for funding from the European Union's Horizon 2020 research and innovation programme under the Marie Sk\l{}odowska-Curie grant agreement No. 800502. AM acknowledges the support from the Portuguese Strategic Programme UID/FIS/00099/2019 for CENTRA. 
AV and AB acknowledge PREMIALE 2015 MITiC. DB is supported in the form of work contract FCT/MCTES through national funds and by FEDER through COMPETE2020
in connection to these grants: UID/FIS/04434/2019; PTDC/FIS-AST/30389/2017 \& POCI-01-0145-FEDER-030389

The preparation of this work has made extensive use of Topcat \citep{Taylor05}, and of NASA's Astrophysics Data System Bibliographic Services, as well as the open-source Python packages \texttt{Astropy} \citep{Astropy13}, \texttt{NumPy} \citep{VanDerWalt11}, and \texttt{scikit-learn} \citep{scikit-learn}. The figures in this paper were produced with Matplotlib \citep{Hunter07}. Figure~\ref{fig:emceeCornerPlot} was produced with \texttt{corner} \citep{corner}.

\bibliographystyle{aa} 
\linespread{1.5}                
\bibliography{references}

\end{document}